\documentclass{aa}  
\usepackage{graphicx}
\usepackage{multirow}
\usepackage[dvipsnames]{xcolor}
\usepackage{txfonts }
\usepackage{natbib  }
\usepackage{hyperref}
\usepackage{enumitem} 
\usepackage{varwidth}
\def\g292{{G292.0+1.8}}
\def\G127{{G127.1+0.5}}
\def\P1124{{PSR~J1124$–$591}}
\def\PSR4FGL{{4FGL~J1124.7$-$5915}}

\begin{document}

\title{Unlocking the radio-$\gamma$  spectrum of the pulsar wind nebula around PSR~J1124$-$5916 in SNR~\g292
}

%-----------------------------------------------------------------------

\author{A. Lemi\`ere
\inst{1}
%\href{https://orcid.org/0000-0002-6682-7188}{\includegraphics[scale=0.8]{figures/orcid.png}}\thanks{E-mail:alemiere@apc.in2p3.fr}
   \and 
   G. Castelletti\inst{2}
   %\href{https://orcid.org/0009-0002-0134-2064}{\includegraphics[scale=0.8]{figures/orcid.png}}
   \and N. L. Maza\inst{3}
  }

%-----------------------------------------------------------------------
\institute{Université Paris Cité, CNRS, Astroparticule et Cosmologie, F-75013 Paris, France 
\email{alemiere@apc.in2p3.fr}
\and 
Instituto de Astronom\'ia y F\'isica del Espacio (IAFE), CONICET-Universidad de Buenos Aires, Intendente Güiraldes 2160 (C1428EGA), Ciudad Aut\'onoma de Buenos Aires - Argentina
\and 
Observatorio Astron\'omico F\'elix Aguilar (OAFA), FCEFyN -- Universidad Nacional de San Juan, Benavidez 8175 (O), San Juan -  Argentina}

%-----------------------------------------------------------------------
\date{Received 03 July 2025 / accepted 04 October 2025}
%-----------------------------------------------------------------------
\abstract
{We present the first detection of GeV $\gamma$-ray emission potentially associated with the pulsar wind nebula (PWN) hosted by the young core-collapse supernova remnant G292.0+1.8, based on a detailed time-resolved analysis of \textit{Fermi}-LAT data. By isolating the unpulsed component from the dominant magnetospheric radiation of PSR~J1124$-$5916, we successfully disentangle 
a candidate nebular emission in the GeV range, characterise its morphology and extract its spectrum. 
This identification places G292.0+1.8 among the few systems in which the pulsar and PWN contributions have been spectrally resolved at high energies, offering new insight into their respective emission mechanisms. We characterise the  $\gamma$-ray spectrum of the pulsar and model the broadband spectral energy distribution (SED) of the PWN using radio, X-ray, and GeV data. The emission is well described by a single electron population with two spectral breaks: one intrinsic to the injection spectrum and another produced by synchrotron cooling in a magnetic field of $\sim$15~$\mu$G. Notably, the inferred magnetic field and the low TeV flux of the nebula resemble those of 
3C~58, suggesting that similar low-field environments can arise in young PWNe. The high-energy portion of the SED is now tightly constrained by our GeV detection and existing TeV upper limits. Compared to our model, earlier predictions tend to underpredict the $\gamma$-ray flux, while others that succeed in reproducing the GeV component often overpredict the TeV emission. This mismatch underscores the challenges in modelling particle acceleration and radiation processes in young PWNe and establishes G292.0+1.8 as a valuable benchmark for testing and refining such models.
}

\keywords{gamma-rays: ISM -- pulsars: individual: \object{PSR~J1124$-$5916} -- ISM: supernova remnants 
-- ISM: individual objects: \object{G292.0+1.8} -- 
%\object{G296.5+10.0} -- 
radiation mechanisms: non-thermal -- pulsar wind nebulae}

\titlerunning{Unlocking the radio-$\gamma$  spectrum of the pulsar wind nebula around PSR~J1124$-$5916 in SNR~\g292}
\authorrunning{Lemi\`ere A. et al.}
\maketitle

\section{Introduction}
\label{sec:introduction}
Pulsars (PSRs), the compact remnants of core-collapse supernovae, rank among the most energetic astrophysical sources in the Galaxy. 
Their magnetised, relativistic winds interact with the surrounding medium to form 
synchrotron-emitting structures known as pulsar wind nebulae (PWNe). 
These nebulae display broad-band non-thermal emission from radio to X-rays, shaped by the properties of the central pulsar (e.g.,  spin-down luminosity and age) and by the characteristics of the local environment (see \citep{gaenslerSlane2006} for a review).
At higher energies, many PWNe have also been detected in $\gamma$-rays, from MeV to TeV, primarily through inverse Compton scattering of ambient photon fields \citep{dejager2009PWNe}\citep{FERMI_PWNTeV2013}\citep{2018_HESS_PWN_Pop}. 
These detections confirm the role of PSRs and their nebulae as key contributors to the leptonic component of Galactic cosmic rays (CRs).

Disentangling the emission from the pulsar and its surrounding nebula, however, remains a significant observational challenge across the electromagnetic spectrum.  
In the radio band, the pulsar's coherent emission often overwhelms  the fainter synchrotron radiation from the PWN. In high energy $\gamma$-rays, the magnetospheric emission from the pulsar can outshine the nebular component--especially below $\sim$10~GeV--while the limited angular resolution of current $\gamma$-ray instruments further complicates the spatial analysis. These difficulties are compounded when the PWN is embedded within its host supernova remnant (SNR), as frequently occurs during the early evolutionary stages. 
As a result, only in a few well‑studied systems, such as the Vela complex, have allowed for a robust isolation of the pulsar and the nebular contribution  through extensive multiwavelength, morphological, time-resolved analyses \citep[e.g.,][]{grondin2013,lange2025}. 
 This paper presents one additional case to that limited set. 
We focus here on resolving the $\gamma$-ray emission components of PSR~J1124$-$5916 and its associated PWN. The pulsar is hosted by the young core-collapse SNR G292.0+1.8, an oxygen-rich remnant estimated to be $\sim$3000 years old \citep{winkler2009}. The remnant exhibits bright thermal X-ray knots and filaments confined within a faint radio shell approximately $8^{\prime}$ in diameter. Its X-ray morphology reveals a complex interplay between shocked circumstellar material (CSM) and metal-rich ejecta (see \citealt{bhalerao2019} and references therein).
A lower limit on its distance  of  $6.2\pm0.9$~kpc  was inferred from  H\,{\sc i} absorption measurements, implying a physical  diameter  of approximately 14.4~pc \citep{gaensler2003}. 
Although no definitive interaction with molecular gas has been confirmed, \citet{lee2010} proposed that variations in electron temperature and emission measure in X-rays may indicate interaction with the red supergiant wind of the progenitor star.

With a  spin-down luminosity of $\dot{E}\sim1.2\times10^{37}$~erg~s$^{-1}$ and a rotation period of 135~ms,  PSR~J1124$-$5916 is among the most energetic known Galactic pulsars \citep{Camilo2002G292}. 
Its wind powers a PWN that extends 
$\sim$$1^{\prime}\!\!.8 \times 1^{\prime}$ in X-rays and $\sim$$4^{\prime}$ in the radio band \citep{hughes2001,gaensler2003}, exhibiting a toroidal morphology with a jet-like structure reminiscent of other young PWNe such as the Crab Nebula and 3C~58 \citep{park2007}. 
This torus has also been detected at optical and near-infrared (NIR) wavelengths  \citep{zharikov2013}. 
% detection in the gamma-ray domain

PSR~J1124$-$5916 was detected  by the \it Fermi \rm Large Area Telescope (\it Fermi\rm-LAT; hereafter LAT) as a bright  pulsed $\gamma$-ray source \citep{Abdo2010,smith2023}.
The LAT spectral analysis reveals a hard power-law with a super-exponential cut-off around a few GeV \citep{smith2023}. However, if the nebular contribution is not properly accounted for, the resulting pulsar spectrum may be biased. Identifying the unpulsed PWN component is therefore essential to accurately characterise the magnetospheric emission and to evaluate the nebular contribution at GeV energies. 
At TeV energies, although the region has been surveyed by the H.E.S.S. Galactic Plane Survey, no detection was achieved; only a an upper limit of $0.27 \times 10^{-12}$~cm$^{-2}$~s$^{-1}$ has been established \citep{2018_HESS_PWN_Pop}.

The structure of this paper is as follows. In Sect.~\ref{sec:2radio-data}, we update   
the radio continuum spectrum of the PWN powered by PSR~J1124$-$5916 by incorporating new data at  
frequencies below $\sim$1~GHz. In Sect.~\ref{sec:3GeVemission}, we analyse  the $\gamma$-ray 
observations from LAT and describe the methodology adopted to disentangle the nebular $\gamma$-ray emission from the pulsed component. Sect.~\ref{sec:4discussion} presents a broadband spectral energy distribution (SED) analysis of the PWN, modelling its emission from radio to $\gamma$-rays and investigating the underlying particle populations and energy loss mechanisms. This section also includes an in-depth discussion of the implications of our results.
Finally, Sect.~\ref{sec:5conclusions} summarises the main conclusions of the study. %our main findings.
\section{The radio dataset}
\label{sec:2radio-data}
\begin{figure*}[h!]
\centering
  \includegraphics[scale=0.5]{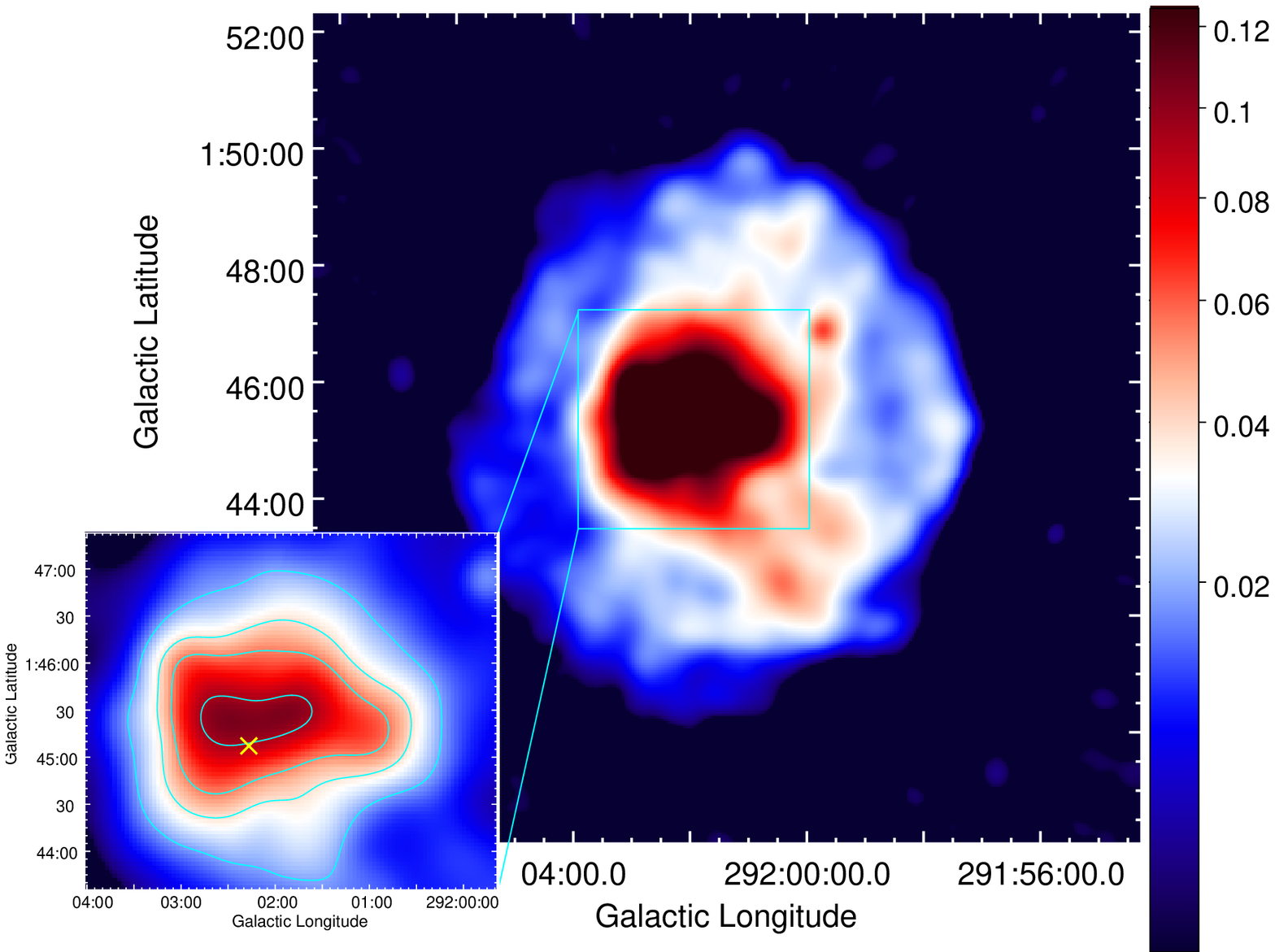}
    \includegraphics[scale=0.67]{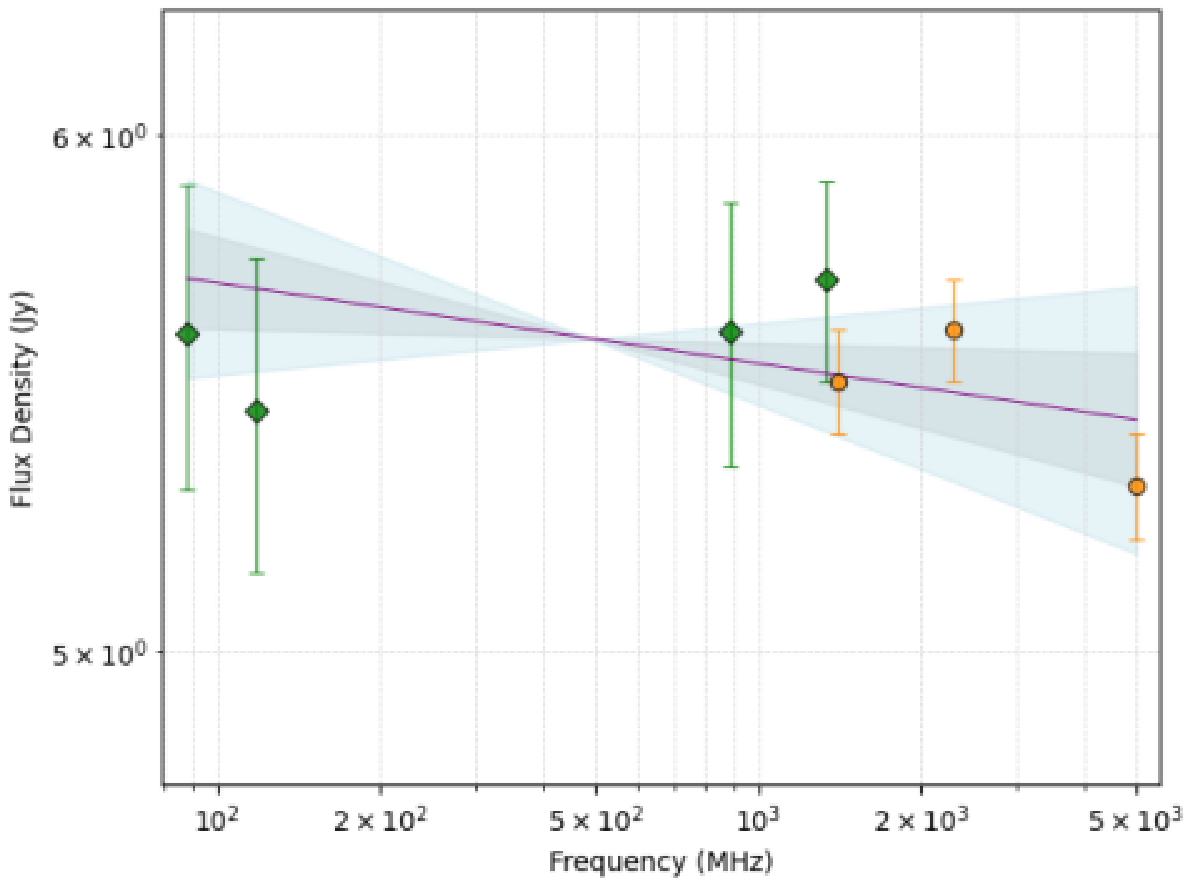}
   \caption{\textit{Left}:  Radio image of the SNR~G292.0+1.8 from the Rapid ASKAP Continuum Survey at 887.5~MHz, showing the full extent of the remnant. The colour scale (square-root scale, in units of Jy~beam$^{-1}$) is adjusted to enhance the visibility of the shell structure; as a result, the central PWN appears saturated. A zoom-in on the PWN is shown in the inset, where the square-root colour scale ranges from 0.015 to 0.4~Jy~beam$^{-1}$ and is optimised to highlight the internal PWN morphology. The contours in the inset are drawn at 0.07, 0.13, 0.18, and 0.3~Jy~beam$^{-1}$ and are shown for reference to delineate internal structures. The yellow cross marks the position of the PSR~J1124$-$5916. The synthesised beam is $25^{\prime\prime} \times 25^{\prime\prime}$, and the rms noise level is 0.5~mJy~beam$^{-1}$. 
   \textit{Right}: Updated radio continuum spectrum of the PWN associated with PSR~J1124$-$5916 within SNR~\g292. Green diamond symbols correspond to the  new flux density measurements obtained in this work from radio surveys, while orange symbols represent values compiled from the literature (see Table~\ref{tab:PWNradiofluxes}).  
 The solid line indicates the best fit to the weighted data, yielding a radio spectral index $\alpha=-0.012\pm0.010$.
 The grey and light blue shaded bands denote the 1$\sigma$ and 2$\sigma$ statistical uncertainties in the best-fit $\alpha$ value, respectively.   }
 \label{fig:PWNradio-spectrum}
\end{figure*}

To anchor the low-energy segment  of the broadband SED of the PWN associated with  PSR~J1124$-$5916, we measured new radio flux densities from the Galactic and Extragalactic All-sky MWA Survey (GLEAM; \citealt{hurley-walker2017}), the Rapid ASKAP Continuum Survey (RACS; \citealt{mcconnell2020}), and MeerKAT L-Band observations \citep{cotton2024}.\footnote{We note that \citet{cotton2024} report a flux density of $5.8 \pm 0.4$~Jy from the same MeerKAT image used here. Our independent measurement (see Table~\ref{tab:PWNradiofluxes}) is fully consistent with this result, and we adopt it for the spectral analysis.} 
Instead of applying fixed geometric regions uniformly across all the radio images, we estimated the contribution from the surrounding SNR shell and the underlying Galactic background individually for each observing frequency. 
This was achieved by inspecting multiple horizontal and vertical intensity slices across the PWN region. These profiles allowed us to assess the combined level of large-scale emission projected along the line of sight--both from the SNR shell (in front of and behind the PWN) and from the diffuse Galactic background. 
The average of this total contribution was then subtracted from the 
flux measured within a region encompassing the PWN.\footnote{A similar procedure was applied to the PWN in IC~443; \citet{swartz2015}.} 
For instance, at 887.5~MHz, the combined  emission unrelated to the PWN amounts to roughly 10\% of the total flux within the selected region. The boundaries of this region, as well as those used to characterise the surrounding SNR emission, were determined by averaging several slightly different estimates of the outer perimeter, thereby accounting for uncertainties in separating the PWN from the extended SNR. A radio image of SNR~G292.0+1.8 is shown in Fig.~\ref{fig:PWNradio-spectrum}(\textit{Left}), including a zoomed-in view of the central region to illustrate the morphology of the PWN and its surrounding structure.  
Using this procedure, our measurements improve the spectral coverage of the PWN at low and intermediate radio frequencies, extending well below the range probed by the earlier analysis of PSRJ1124$-$5916's PWN presented by \citet{gaensler2003}.

Table~\ref{tab:PWNradiofluxes} summarises the flux densities employed in our analysis. 
To ensure a uniform flux scale, the new measurements were corrected to the absolute calibration of \citet{perley17}. 
Flux densities sourced from \citet{gaensler2003} could not be corrected, as the calibrator used in their observations is not included in the Perley \& Butler flux density scale. However, the expected correction factor is close to unity, and thus any systematic discrepancy is negligible within the uncertainties. 

The compiled flux densities were then fitted with a power-law model of the form $S_\nu \propto \nu^\alpha$, where $S_\nu$ is the flux density at frequency $\nu$, and $\alpha$ is the spectral index. A single power-law slope  $\alpha = -0.012 \pm 0.010$ provides an excellent fit to the data with a reduced $\chi^2_\mathrm{\nu} = 0.95$ (see Fig.~\ref{fig:PWNradio-spectrum}, \textit{Right}). This result is slightly flatter than the spectrum obtained when fitting only the high-frequency measurements ($\alpha_{1400-5000\,\mathrm{MHz}} \sim -0.03$, $\chi^2_\mathrm{\nu} = 2.1$; \citealt{gaensler2003}), but remains consistent with the characteristically flat radio spectra observed in many PWNe, which typically exhibit spectral indices between 0.0 and $-0.3$. Such flat spectra are generally attributed to the continuous injection of relativistic electrons by the pulsar wind, coupled with efficient synchrotron emission within the nebular magnetic field \citep{gaenslerSlane2006}. The complete radio dataset and the corresponding model will be included in a forthcoming catalogue of SNR radio continuum spectra currently in preparation (Castelletti et al. in prep). 

\begin{table}[h!]
\small\centering
\caption{Integrated flux densities compiled for the radio PWN around PSR~J1124$-$5916, used to construct the spectrum shown in Fig.~\ref{fig:PWNradio-spectrum}. }
%-----------------------------------------------------------------
\begin{tabular}{lcl}
\hline\hline
%------------------------------------------------------------------
  Frequency    &  Flux density     &  Reference  \\
   MHz  &     Jy       &           \\ 
    \hline
    88  &	$5.6 \pm 0.3$ &  This work (GLEAM)  \\ %\tablefootmark{~$\ast$} 
    118 &   $5.4 \pm 0.3$ &  This work  (GLEAM) \\
    887.5 & $5.6 \pm 0.3$ &  This work  (RACS) \\
    1,335 & $5.7 \pm 0.2$ & This work   (MeerKAT) \\
    1,400 &$5.5 \pm 0.1$ & \citet{gaensler2003} \\ 
    2,300 & $5.6 \pm 0.1$ & \citet{gaensler2003} \\
    5,000 &$5.3 \pm 0.1$ & \citet{gaensler2003} 
    \\\hline
 \end{tabular}
\label{tab:PWNradiofluxes}
\end{table}

\section{LAT data analysis}
\label{sec:3GeVemission}
The LAT, aboard the \it Fermi \rm Gamma-ray Space Telescope, is a pair-conversion $\gamma$-ray detector designed to survey the sky from 20~MeV to 1~TeV \citep{atwood2009}. 
The third catalogue of LAT-detected pulsars (3PC) currently lists over 340 bright point sources exhibiting pulsed $\gamma$-ray emission \citep{smith2023}. 
Many of these pulsars are expected to power associated  
PWNe, as a significant fraction of their spin-down energy is channelled 
into a magnetised particle wind. 
This wind interacts with the surrounding supernova ejecta, continuously feeding the nebula and sustaining its non-thermal emission, which can potentially be detected in the LAT energy range. However, below $\sim$10~GeV, the pulsed $\gamma$-ray emission from the pulsar typically dominates the signal, complicating the identification and characterisation of any underlying PWN contribution in LAT data \citep{ackermann2011fermiPWN,2012PWNJ1857}.
PSR~J1124$-$5916 (associated with the source \PSR4FGL)  
has been identified as a bright pulsed emitter in all \it Fermi \rm LAT pulsar catalogues to date \citep{Abdo2010,Abdo_2013,smith2023}. 
Its phase-averaged $\gamma$-ray spectrum displays strong emission up to $\sim$10~GeV, beyond which a pronounced spectral cut-off is observed. Intriguingly, above this cut-off, a residual emission component persists, extending up to $\sim$1~TeV \citep{smith2023}. In this section, we investigate the possibility that this high-energy residual emission originates from a yet undetected PWN, by providing a detailed data analysis of the LAT available dataset. 
\subsection{Analysis framework}
\label{sec:2.2general-gamma-data}
\subsubsection{Data reduction}
\label{sec:2.2.1gamma-selec-reduc}
Data reduction and exposure calculations were carried out  using the \it Fermi \rm Science Tools (\it fermitools\rm, version 2.2.0) in conjunction with the  \it fermipy \rm (version 1.1.6) analysis package \citep{Wood+17_fermipy}. 
We employed the Pass 8 third release (P8R3) photon data 
and the corresponding  instrument response functions (IRFs: P8R3\_SOURCE\_V6), which offer improved event reconstruction and enhanced background rejection \citep{atwood2013, Ajello_2021}. 

To ensure data quality and minimise systematic uncertainties, we excluded events with energies below 50~MeV, where the LAT response function becomes increasingly uncertain. 
We further applied a zenith angle cut of $< 90^{\circ}$ to reduce contamination from Earth limb $\gamma$-rays, and restricted the analysis to good time intervals, 
thus excluding periods affected by transient spacecraft conditions or South Atlantic Anomaly passages, where increased radiation levels can degrade instrument performance.

\subsubsection{Binned likelihood analysis}
\label{sec:2.2.2likelihood}
Full dataset $\gamma$-ray analysis was performed over the 50~MeV to 1~TeV energy range using logarithmic energy binning, with eight bins per decade, across a $6^{\circ}\times6^{\circ}$ ROI centred on the position of \PSR4FGL. 
A standard binned maximum-likelihood analysis was conducted using a source model that incorporates  both diffuse emission components and nearby $\gamma$-ray sources from the incremental Fourth \it Fermi\rm-LAT Source Catalogue (4FGL-DR4; \citealp{abbollahi2022,Ballet+23})  located within 10$^{\circ}$ of the target position.  
To improve the robustness of the analysis, sources with a test statistic TS $<$  4 ($\sigma$$<$2) were excluded from the model. 
The Galactic diffuse emission was modelled using the  {\it gll\_iem\_v06} template, while the isotropic background was described with the  {\it iso\_P8R2\_SOURCE\_V6\_v06} model.\footnote{\url{https://fermi.gsfc.nasa.gov/ssc/data/access/lat/BackgroundModels.html}} The scaling factor for the Galactic diffuse component and the normalisation of the isotropic background were both left free to vary during the fit. \\
Although the LAT point spread function (PSF) degrades at low energies, 
we opted to perform both the morphological and spectral analyses over the full energy range. 
This strategy maximizes sensitivity and improves our ability to disentangle the contribution from the pulsar itself and any extended emission potentially associated with the surrounding nebula.
As part of the subsequent analysis, we produced test statistic (TS) maps for $\gamma$-ray photons with energies exceeding~1 GeV, in order to examine the morphology of the source in an energy range where the point spread function (PSF) is sufficiently well constrained to allow meaningful spatial characterisation.

\subsubsection{Systematic errors evaluation}
\label{sec:2.2.3errors}
A key source of systematic uncertainty in the spectral analysis arises from limitations in the instrument response function, particularly in the estimation of the LAT effective area. To account for this, we applied two bracketing scaling functions to the effective area, following the methodology outlined in \citet{Ackermann+12_Fermi}.

For sources located above the Galactic plane, additional systematic uncertainties in the $\gamma$-ray spectrum arise from the modelling of large-scale emission components.
These components include contributions from extended 
nearby sources as well as the diffuse Galactic $\gamma$-ray emission (DGE).
To assess the impact of these uncertainties, we varied the normalisation of the DGE model by $\pm6$\%, following the approach adopted by 
\citet{Abdo2009_W51C} and consistent with the level of systematics evaluated by \citet{ackermann2012_DGE}.
The two dominant sources of systematic uncertainty -- instrument response and diffuse emission modelling -- were then combined in quadrature with the statistical errors to determine the total uncertainty intervals associated with the flux points presented in the $\gamma$-ray spectrum.

\subsection{Phase resolved analysis}
\label{sec:3.1phase-analysis}
To explore the possibility that part of the high-energy emission in the direction of \PSR4FGL originates from a yet undetected PWN, we first performed a phase-resolved spectral analysis. This approach involves isolating the unpulsed component by filtering out the pulsed signal from the dataset, thereby enhancing sensitivity to any steady $\gamma$-ray contribution from the nebula.

\subsubsection{Data selection and preparation}
The phase-resolved analysis requires 
assigning rotational phases to $\gamma$-ray photons and selecting %only those 
those that fall within an OFF-pulse window. 
These phased events were obtained from the FT1 event file provided by the 3PC, 
made available through the FSSC data access portal. 
\footnote{(\url{https://fermi.gsfc.nasa.gov/ssc/data/access/lat/})} 
%\footnote{(\url{https://fermi.gsfc.nasa.gov/ssc/data/access/lat/3rd_PSR_catalog/3PC_HTML/J1124-5916.html})}  
The file includes all reconstructed events in the 50~MeV--300~GeV energy range in a $3^{\circ}$ region around PSR~J1124$-$5916  and features  two additional columns: PHASE and MODEL$\_$WEIGHT. These columns contain, respectively,  the rotational phase of each photon and the probability that the photons originate from the pulsar, based on LAT's energy-dependent PSF and the angular offset between the photon direction and the pulsar position.
Photon arrival times were converted into rotational phases using pulsar timing models derived 
long-term timing campaigns, as described in the 3PC \citep{smith2023}.
\begin{figure}[h!]
    \includegraphics[scale=0.6]{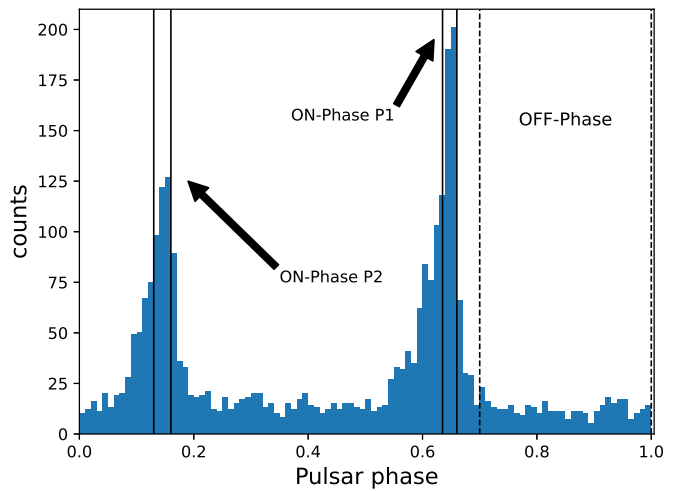}
 \caption{Phase profile constructed using photons with energies above 50~MeV in a region of $0^{\circ}\!\!.2$ around PSR~J1124$-$5916. 
  The profile spans one full rotation, divided into 100 bins.
 The two dashed and solid vertical lines represent the off-pulse and on-pulse windows, respectively, as defined in the text and used in the phase resolved analysis.}
 \label{fig:Pulsar_phase}
\end{figure}
Figure~\ref{fig:Pulsar_phase} displays the phase-folded pulse profile of PSR~J1124$-$5916, 
constructed from photons with energies  
$E >$ 50~MeV within an angular radius of $0.^{\circ}\!\!2$   
centred on the pulsar. 
Two distinct peaks, previously identified  by \citet{Abdo2010,Abdo_2013} and \citet{smith2023}, are clearly visible and separated by $\sim0^{\circ}\!\!.5$ in phase. In the following, we refer to the brighter peak, centred at phase $0^{\circ}\!\!.65$, as P1, and to the smaller peak, centred at $0^{\circ}\!\!.16$, as P2. 

The OFF-pulse regions of the light curve, located between P2 and P1 (phase intervals 0.2--0.5 and 0.7--1.0),  span over 50\%  of the pulsar's rotational period. 
For the purposes of this study, we defined the ON-pulse intervals as 0.635--0.66 for P1 and 0.13--0.16 for P2, consistent with prior \it Fermi\rm-LAT analyses \citep{Abdo2010}. 
These ON-pulse windows were defined by restricting 
to regions where the photon count exceeded 50\% of the maximum peak amplitude, in order to minimise contamination from potential PWN emission.  
To ensure the robustness of our analysis and to minimise contamination from potential residual pulsed emission between peaks, we adopted the OFF-pulse phase interval 0.7--1.0, consistent with the definition used in the \it Fermi\rm-LAT pulsar catalogue \citep{Abdo2010}. 
The full dataset was thus segmented into three phase intervals -- P1, P2, and OFF -- and each window was analysed independently using the binned likelihood method outlined in Sect.~\ref{sec:2.2.2likelihood}.

To prevent contamination of OFF-pulse phase regions caused by model inaccuracies at later epochs, 
we restricted the phase-resolved analysis to the valid timing intervals specified in the 3PC. 
This interval spans 11.8 years of LAT observations, from MJD~54682 (August 4, 2008) to MJD~59000 (May 31, 2020) \citep{smith2023}.

\subsubsection{OFF-phase analysis: Search for unpulsed emission}
\label{sec:3.1.1off-analysis}
\begin{figure}[h!]
    \includegraphics[scale=0.65]{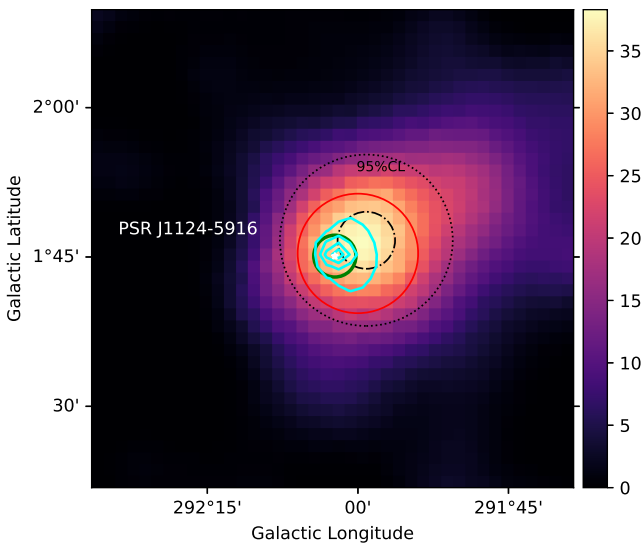}
 \caption{\textit{Fermi}-LAT TS map of the region surrounding
 PSR~J1124$-$5916 for energies above 1~GeV in Galactic coordinates during the OFF-pulse phase interval. All sources of the 4FGL-DR3 catalogue were included, except the pulsar source \PSR4FGL which was removed from the model. 
 The colour bar indicates the TS value range. The white star marks the position of PSR~J1124$-$5916. The black dash-dot circle represents the 1$\sigma$ error radius for the best-fit position obtained in the OFF-pulse phase interval analysis, while the black dashed circle indicates the 95\% confidence level upper limit on the source's extension. Light blue contours show radio data from MOST highlighting the extent of the radio PWN, the green circle correspond to the X-ray PWN extension detected by \it Chandra \rm \citep{hughes2003},  and the red circle indicate the SNR G292.0+1.8 extension.}
 \label{fig:PWN_OFF_TSMAP}
\end{figure}

The OFF-pulse dataset was analysed using the  likelihood adjustment method described in Sect.~\ref{sec:2.2.2likelihood}. Initially, all sources within 15$^{\circ}$ of PSR~J1124$-$5916's position listed in the third incremental update to the original 4FGL catalogue (i.e., 4FGL-DR3) 
were included in the source model, adopting their spectral forms and parameters as given in the catalogue.  
For sources within a 3$^{\circ}$ radius of PSR~J1124$-$5916, the spectral normalisation parameters were allowed to vary freely, while all other source parameters were fixed to their catalogue values. 
The normalisations of both the Galactic and extragalactic diffuse emission components were also left free during the fit. 
To search for residual unpulsed emission, we removed \PSR4FGL (the point source associated with PSR~J1124$-$5916) from the source model and generated a residual TS map.
As shown in Fig.~\ref{fig:PWN_OFF_TSMAP}, the TS map revealed significant excess 
$\gamma$-ray emission near the pulsar position,  
with a peak TS value of $\sim$37, corresponding to a detection significance of approximately 6$\sigma$.

We then introduced a new point source at the pulsar's position, modelled with a simple power-law spectrum. A likelihood fit was performed with all source parameters free to vary.
The inclusion of this source resulted in a $\Delta$TS= 37, confirming a detection at the 6.1$\sigma$ level. 
Using \it fermipy\rm, we localised the new source to Galactic coordinates 
$l = 291^{\circ}\!\!.99 \pm 0^{\circ}\!\!.03$, 
$b = 1^{\circ}\!\!.77 \pm 0^{\circ}\!\!.03$, and determined a 68\% confidence positional uncertainty of $0^{\circ}\!\!.047$ (see Table~\ref{tab:Positions_Pulsar_table}).
To evaluate whether the emission was spatially extended, we fitted a two-dimensional symmetric Gaussian model using the \textit{gta.extension} function in \it fermipy\rm. 
The extension significance, calculated using the test statistic TS = 2 $\Delta \log$ (likelihood), yielded TS = 0.6. 
This result indicates that the emission is not significantly extended, consistent with an upper limit on the extension at the 95\% confidence level of $0^{\circ}\!\!.143$. 
Figure~\ref{fig:PWN_OFF_TSMAP} displays the TS map above 1~GeV, overlaid with the 68\%  confidence error radius for the best localisation and the 95\% confidence upper limit on the source extension.

The $\gamma$-ray spectrum of the candidate source 
was extracted assuming a point source located at the best-fit position.  
 We performed a maximum likelihood fit across eight logarithmically spaced energy bins per decade, spanning the 50~MeV to 300~GeV range. 
The resulting spectral shape is well represented by a hard power law, characterised by a photon index  $\Gamma = 1.99 \pm 0.15$ and a total energy flux 
$6.399 \times 10^{-12}$~erg~s$^{-1}$~cm$^{-2}$ above 50~MeV.

The absence of a significant spectral cut-off up to 300~GeV in the power-law spectrum, coupled with the relatively hard photon index, strongly suggests that this emission does not originate from residual magnetospheric radiation in the OFF-pulse window of PSR~J1124$-$5916.
The spatial coincidence of the detected source with the radio and X-ray PWN of 
PSR~J1124$-$5916 supports this association. In contrast, the radio emission from the shell of G292.0+1.8 exhibits a plateau-like morphology, with an approximately constant flux that extends well beyond the  
observed size of the $\gamma$-ray source, out to a radius of $\sim$4$^{\prime}$.  
Furthermore, the lack of detected synchrotron X-ray emission from the shell challenges the scenario in which the $\gamma$-ray emission arises from high-energy electrons associated with the SNR shell. 
Although a contribution from hadronic processes--i.e., accelerated protons in the shell interacting with ambient material--cannot be entirely ruled out, 
we consider this explanation less likely given the moderate ambient density (0.5–0.9 cm$^{-3}$) inferred from previous studies \citep{gaensler2003, Lee2009}.

\subsubsection{ON-phase analysis: Measurement of PSR~J1124$-$591 pulsed spectrum}
\label{sec:3.1.2on-analysis}
In this section, we analyse the ON-phase $\gamma$-ray data corresponding to the P1 and P2 phase windows. A standard binned likelihood analysis was performed on the LAT data, %as described in Sect.~\ref{sec:2.2.2likelihood}.
following the methodology described in Sect.~\ref{sec:2.2.2likelihood}. 
To account for the contribution from the PWN candidate identified in the OFF-pulse analysis, we incorporated its emission into the ON-pulse fit. Specifically, the PWN candidate emission was included as a fixed component in the spectral model for the pulsed emission analysis with its contribution adjusted by rescaling the exposure to reflect the shorter time interval of the ON-pulse phase. 
The contribution of the pulsar PSR~J1124$-$5916 (\PSR4FGL) was modelled using a power law with a super-exponential cut-off, known as the PLEC4 model (PLSuperExpCutoff4 in the \it Fermitools\rm). 
This spectral form, which includes a super-exponential index $b < 0$, is consistent with that used in the 4FGL-DR3 catalogue \citep{abbollahi2022} and the 3PC \citep{smith2023}. 
The functional form of the PLEC4 model is given by: 
\begin{equation}
\frac{dN}{dE} = N_{0} \, \left(\frac{E}{E_0}\right)^{-\gamma + \frac{d}{b}} 
\exp\left[ \frac{d}{b^2} \left( 1 - \left(\frac{E}{E_0}\right)^b \right) \right],
\end{equation}
\noindent
where $N_{0}$ is the normalisation, 
$\gamma$ is the power-law index, $d$ is the local curvature at the reference energy $E_0$ (in GeV), and $b$ is the super-exponential index, which was fixed to 2$/$3. 
The best-fit spectral parameters for the pulsed emission of PSR~J1124$-$5916 are summarised in Table~\ref{tab:spectra_Pulsar_table}. The resulting energy flux, spectral index, and exponential cut-off parameters are consistent with values reported in the 4FGL-DR3 and 3PC 
catalogues \citep{abbollahi2022,smith2023}. 
\begin{figure}[h!]
    \includegraphics[scale=0.45]{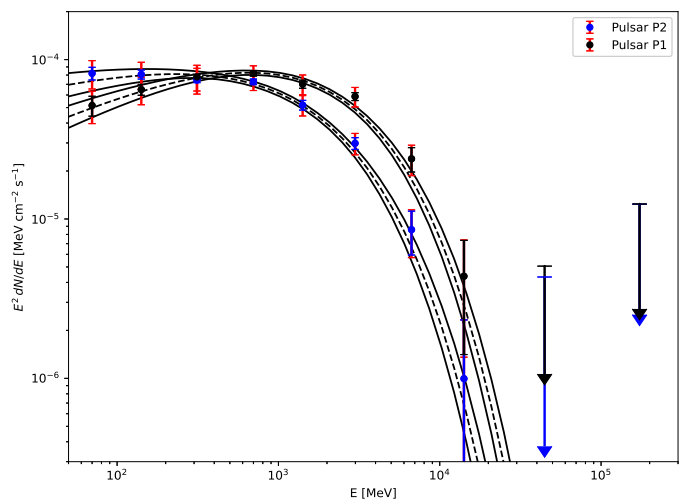}
 \caption{\textit{Fermi}-LAT $\gamma$-ray spectra of the pulsar PSR~J1124$-$5916 (\PSR4FGL) for the pulsed P1 peak (black dots) and pulsed P2 peak (blues squares) phase intervals. The solid black and  dashed blue curves represent the best-fit 
 sub-exponentially cut-off power-law models for the P2 and P1 phase intervals, respectively, over the  50~MeV$-$300~GeV energy range. The model shape follows the one used for significantly curved pulsars in the 4FGL-DR3 catalogue. Red uncertainties account for both statistical and systematic contributions, with the latter primarily arising from the Galactic diffuse emission model and IRFs at low energy. The butterfly region denotes the 1$\sigma$ confidence interval of the best-fit spectral model.}
 \label{fig:P1_P2_Spectrum}
\end{figure}

Figure~\ref{fig:P1_P2_Spectrum} shows the best-fit spectral model along with the flux points derived for the pulsed emission of PSR~J1124$-$5916. 
The flux points were obtained by dividing the energy range from 50~MeV to 300~GeV into 10 logarithmically spaced energy bins. 
In each bin, a maximum likelihood spectral analysis was performed using a power-law shape with a fixed photon index of 2 for the source. 
This was accomplished using the \textit{gta.sed} function in \it fermipy\rm.
During the bin-by-bin analysis, the normalisations of the diffuse Galactic and isotropic emission components, as well as those of all sources within 3$^{\circ}$  of the pulsar position, were allowed to vary freely.  
A 95\% confidence upper limit was calculated when the significance value of a flux point was below 3$\sigma$. 
\begin{table}[h!]
\small
\caption{LAT morphological parameters of the new PWN candidate associated with PSR~J1124$-$5916.}
\label{tab:Positions_Pulsar_table}
%---------------------------------------------------------------------------
\begin{tabular}{l c c c }\hline\hline
   Phase       & $l$       & $b$ & UL Size  95CL  \\
               &  deg.             & deg.                &   deg.             \\\hline
%---------------------------------------------------------------------------
   OFF-pulse      & $ 291.99 \pm 0.03 $   &  $1.78 \pm 0.03$   &   $0.143$        \\
   TOT Dataset     & $ 292.00 \pm 0.03  $  &  $ 1.73 \pm 0.03 $   &   $0.165$       \\\hline
\end{tabular}
\tablefoot{The morphological parameters have been derived above 50~MeV from the OFF pulse analysis and the Full dataset analysis. 
 The Galactic coordinates of the source centroid obtained in each analysis are reported with statistical uncertainties, along with 95\% confidence level upper limits on the source extension.}
\end{table}
\subsection{Analysis with the full LAT dataset}
\label{sec:3.2morphology-spectrum}
To improve statistical reliability and derive the most accurate spectral and morphological parameters for the newly detected source, we performed a final analysis using the entire LAT dataset available. Since the timing analysis used in the previous section does not cover the full observation period considered here, and given that PSR~J1124$-$5916 is known to undergo frequent glitches \citep{Ge_2020}, our analysis of the full dataset is not phase-resolved. Instead, we rely on the distinct spectral shapes of \PSR4FGL and the detected PWN candidate, which, combined with the large photon statistics, allow us to separate the two sources.

\subsubsection{Data Selection}
For this study, we analysed a region of interest (ROI) centred on the 
position of PSR~J1124$-$5916, corresponding to the 4FGL catalogue source \PSR4FGL (R.A.=11h24m29.57905s, Dec.=$-$59d12m49.9589s).
The ROI spans a $6^{\circ}\times6^{\circ}$ area, and the dataset comprises  15 years of LAT observations, from August 4, 2008 to August 5, 2023. This extensive time baseline provides a high-statistics dataset suitable for detailed spectral modelling, enabling a refined investigation of the $\gamma$-ray emission that was unfeasible in prior studies \citep{smith2023}.

\subsubsection{Morphological study of the PWN candidate}
For the morphological analysis, we adopted the same binned likelihood fitting procedure outlined in Sect~\ref{sec:2.2.2likelihood}. 
The source model incorporated both the known $\gamma$-ray pulsar \PSR4FGL and the newly detected PWN candidate, along with all other catalogue sources within a 4$^{\circ}$  radius of \PSR4FGL, as listed in the 4FGL-DR3 catalogue.
Sources within a 4$^{\circ}$ radius of PSR~J1124$-$5916 had their spectral normalisation parameters allowed to vary freely, while all other source parameters were fixed to their catalogue values. Additionally, the normalization factors of both the Galactic and extragalactic diffuse emission components were allowed to vary during the fitting process.
\begin{table*}
\small
\centering
\label{tab:spectra_Pulsar_table}
\caption{Spectral results for the $\gamma$-ray pulsar PSR~J1124$-$5916.}
\begin{tabular}{l c c c c c c }\hline\hline
   Source    &   Phase         & {$\Gamma$}  & {$d$}      & {$b$}   & { Energy Flux ($>$50~MeV)} & { TS} \\
                      &             &   &         &  & { (10$^{-11}$~erg~cm$^{-2}$~s$^{-1}$)  }      &       \\\hline
 PSR~J1124$-$5916   &    ON P1        & $ -2.134  \pm  0.037 $  & $ 0.406 \pm 0.048 $ & $ 0.66 $ & $  50.1 \pm 2.3 $    &  $3519 $ \\
 PSR~J1124$-$5916   &    ON P2        & $ -2.421  \pm 0.041 $  & $0.461 \pm 0.068$& $ 0.66 $ & $  46.5 \pm$ 2.7     &  $2172 $ \\
 PSR~J1124$-$5916   &  TOT Dataset    & $-2.287 \pm 0.021   $   & $0.38 \pm 0.03$   & $ 0.66 $ & $  6.946  \pm 0.186$ & $ 6054$ \\\hline
 PWN   &    OFF             & $ -1.991 \pm 0.145 $    &                     &                  & $ 0.79  \pm 0.18$ &  $38 $  \\
PWN   &   TOT Dataset      & $ -2.061 \pm 0.077  $ &                     &                  & $  0.66 \pm 0.01 $  & $ 49$ \\\hline
\end{tabular}
\tablefoot{Maximum likelihood spectral fit results for the $\gamma$-ray pulsar PSR~J1124$-$5916 in the ON-phase and average phase, and for our new source associated to the PWN candidate hosted by \g292, in the OFF-phase and average phase total dataset analysis. The pulsar spectrum is modelled using a subexponentially cut-off power-law, consistent with the approach applied to significantly curved pulsars in the 4FGL-DR3 catalogue \citep{4FGLDR32022}. The integral energy flux for E$>$50~MeV, the  low-energy photon index ($\Gamma$), the exponential factor ($d$), and the exponential index ($b$) are provided for each fit pf the pulsar. For the PWN, the spectrum is well described by a simple power-law model. The integral energy flux for E$>$50~MeV along with the photon index $\Gamma$ are reported. The phased analysis (ON P1, ON P2, and OFF) is carried out over the range 50~MeV–300~GeV, whereas the full dataset analysis extends over  50~MeV–1~TeV.}
\end{table*}
\begin{figure*}[t!]
    \includegraphics[scale=0.5]{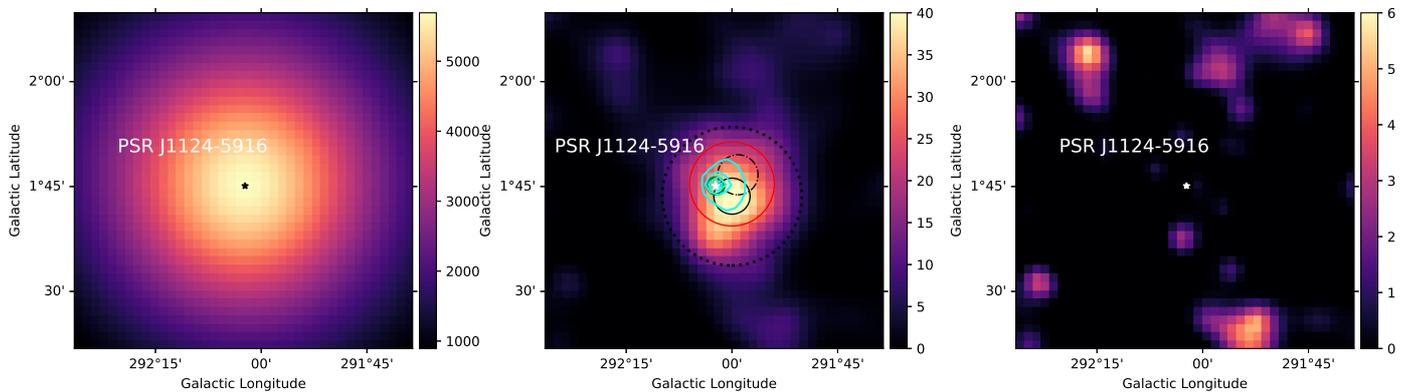}
 \caption{Zoomed-in views of the $>$1~GeV \textit{Fermi}-LAT total dataset TS maps in a  $2^{\circ} \times 2^{\circ}$ region centred on \PSR4FGL. 
 \it Left\rm: Residual TS map after including all sources from the 4FGL-DR3 catalogue except the pulsar PSR~J1124$-$5916 / \PSR4FGL,  and the new detected source. The map highligh the pulsar emission.
 \it Middle\rm: Residual TS map obtained after modeling all 4FGL-DR3 sources, including \PSR4FGL, with their best-fit parameters derived from the likelihood analysis. The map highlights the excess emission associated with a newly detected source associated with the PWN hosted in SNR~G292.0+1.8. The black dash-dotted circle indicates the best-fit position of the new source, with the radius corresponding to the 68\% confidence positional uncertainty. The 95\% confidence upper limit on the source extension is represented by a larger black dash-dotted circle. The position of PSR~J1124$-$5916 / \PSR4FGL is marked with a white star. Radio contours are overlaid in cyan, the green circle corresponds to the X-ray PWN extension detected by \it Chandra \rm \citep{hughes2003},  and the red circle indicates the SNR G292.0+1.8 extension.
 \it Right\rm: Residual TS map for the complete source model, including all 4FGL-DR3 sources, PSR~J1124$-$5916 /\PSR4FGL, and the new detected source, demonstrating that the emission is well accounted for.
 }
\label{fig:TOT_TSMAP}
 \end{figure*}

In this analysis, the PWN candidate is detected with a statistical significance of 7$\sigma$, while \PSR4FGL reaches an impressive 95$\sigma$ detection level. The best-fit position and spatial extension of the PWN candidate, updated using the full LAT dataset, are reported in Table~\ref{tab:Positions_Pulsar_table}. 
The position of the detected PWN candidate is consistent with that obtained from the OFF-pulse analysis, with only a minor offset of a few arcminutes.  
We tested for a possible spatial extension of the source, obtaining a best-fit Gaussian width of 0.$^{\circ}$104 at a significance level of 2.4$\sigma$. Since this does not constitute a significant detection of extension, 
we established a 95\% confidence level upper limit on the source extension of 0.$^{\circ}$165, closely matching the value inferred from the OFF-pulse analysis.
Figure~\ref{fig:TOT_TSMAP} displays, from left to right, the LAT TS maps corresponding to (i) the full model excluding the pulsar \PSR4FGL, (ii) the full model excluding the newly detected PWN candidate, and (iii) the residual excess map from the full model including both sources.  
The maps demonstrate that the distinct spectral characteristics of the pulsar and the PWN enable a clean separation of their contributions. The total model  adequately reproduces the observed $\gamma$-ray emission in the 50~MeV–1~TeV energy range.
As shown in Fig.~\ref{fig:TOT_TSMAP},  the $\gamma$-ray emission associated with the newly detected source coincides spatially with the radio PWN of PSR~J1124$-$5916, which spans a radius of $130^{\prime \prime}$, corresponding to 3.8~pc at a distance of 6~kpc  
\citep{gaensler2003}
The pulsar PSR~J1124$-$5916 also powers an X-ray PWN, consisting of a  compact bright structure of 5$^{\prime\prime}$ and a more extended component reaching out to 65$^{\prime\prime}$ \citep{hughes2001,safiharb2002,hughes2003,park2007}, highlighted by the green circle in Fig.~\ref{fig:TOT_TSMAP}.

Generally the discrepancy in size and morphology between the X-ray and $\gamma$-ray PWNe reflects the distinct populations of electrons responsible for each emission process. The X-ray PWN arises from  synchrotron radiation produced by freshly injected  high-energy electrons, while the $\gamma$-ray PWN is primarily shaped by older, lower-energy electrons undergoing IC scattering \citep{mattana2009, dejager2009PWNe}. In the case of the PWN candidate in SNR~G292.0+1.8, the $\gamma$-ray emission is not spatially resolved; however, the upper limit we derive on its extent exceeds the size of the X-ray PWN. 
The X-ray morphology of the PWN is relatively unaffected by the pulsar's motion or interactions with the SNR's reverse shock, which are expected  in systems a few thousand years old.  In contrast, the $\gamma$-ray PWN morphology is more sensitive to these dynamical effects. 
Given their lower energy, the electrons responsible for radio synchrotron emission are expected to trace the larger-scale structure of the $\gamma$-ray PWN.

In the case of SNR~\g292, PSR~J1124$-$5916  is offset by approximately  $46^{\prime\prime}$ ($\sim$1.34~pc  at a  distance of $\sim$6~kpc) south-east of the presumed explosion centre, implying a transverse velocity of $\sim 500$~km~s$^{-1}$. 
As shown in Fig.~\ref{fig:TOT_TSMAP}, the pulsar is located near the southeastern edge of the radio PWN, roughly $90^{\prime\prime}$ from nebula’s outer edge in the direction of motion, and $\sim 160^{\prime\prime}$  from the opposite boundary.
This pronounced asymmetry suggests that the PWN has evolved within an inhomogeneous ambient medium,  shaped by the pulsar’s motion and potentially influenced by environmental gradients or interactions with the reverse shock.

\subsubsection{Spectrum}
\begin{figure}[h!]
\includegraphics[scale=0.57]{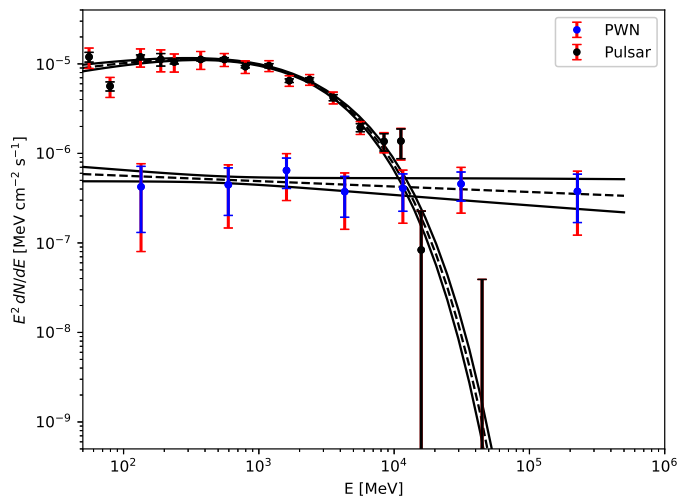}
 \caption{LAT spectrum of the new  $\gamma$-ray PWN candidate associated with \PSR4FGL/PSR~J1124$-$5916, fitted simultaneously over the 50~MeV-1~TeV energy interval using the full dataset. The butterfly region denotes the 1$\sigma$ confidence interval of the best-fit spectral model. Black and blue error bars indicate statistical uncertainties, while red error bars correspond to the total uncertainties, combining statistical and systematic errors in quadrature. Systematic errors do not include uncertainties due to possible sources confusion.}
 \label{fig:TOT_Spectra}
\end{figure}

Using the best-fit spatial parameters derived for the new PWN candidate, we re-performed the spectral analysis over the 50~MeV-1~TeV energy range, 
including all sources from the 4FGL-DR3 catalogue within a $10^{\circ}$ radius, following the same approach as described in previous sections. 
The spectral parameters obtained for \PSR4FGL and the newly detected source are reported in Table~\ref{tab:spectra_Pulsar_table}, along with their associated statistic and systematic uncertainties. The spectrum of the new PWN candidate is well described by a power law with a hard index up to 500~TeV, fully compatible with  the spectral shape obtained from the OFF-pulse analysis. 
To construct the SEDs for both sources, we divided the energy range into ten logarithmically spaced energy bins and derived the corresponding LAT spectral flux points. The resulting SEDs of PSR~J1124$-$5916 and its PWN candidate are shown in Fig.~\ref{fig:TOT_Spectra}. Systematic uncertainties associated with each flux point were estimated using the procedure described in Sect.~\ref{sec:2.2.3errors}. 
Assuming the PWN candidate is associated with PSR~J1124$-$5916, the luminosity of this source is estimated to be 
$L(\mathrm{50~MeV - 1~TeV})=  1.76 \times 10^{34}(D_{6~\mathrm{kpc}})^{2}$~erg~s$^{-1}$, for a distance $D=6$~kpc.
Note that the adopted distance of 6~kpc for the PWN is well-supported by multiple independent measurements, including $6.2 \pm 0.9$~kpc from H\,{\sc i} emission and absorption  and  $6.4 \pm 1.3$~kpc based on the pulsar’s dispersion measure \citep{gaensler2003}.

\section{Discussion}
\label{sec:4discussion}
\subsection{General context}
In the Galaxy, only five other PWNe 
--aside from the one hosted by SNR~G292.0+1.8-- powered by pulsars with spin-down luminosities exceeding $10^{37}$~erg~s$^{-1}$ have been detected in the GeV range with LAT: the Crab Nebula, 3C~58, and those within the SNRs Vela, MSH~15$-$52, and  
HESS~J1813$-$178 \citealp{abbollahi2022,Ballet+23}.  
With the exception of Vela, all these systems are younger than 10~kyr, with two of them --the Crab Nebula and 3C~58-- being very young with estimated ages below 5~kyr \citep{Rudie2007CrabNebulaAge,fesen2008}. 
The LAT detection of the PWN in G292.0+1.8 thus adds a third member to this rare group of very young and powerful sources.

While we demonstrated that the PWN in G292.0+1.8 shines brightly in the GeV range, it stands out at TeV energies as the only non-detected nebula among Galactic systems hosting pulsars with spin-down luminosities above $10^{37}$~erg~s$^{-1}$.  
The upper limit on its TeV luminosity, as reported in the H.E.S.S. collaboration's study of the Galactic PWN population \citep{2018_HESS_PWN_Pop}, implies a maximum TeV luminosity of $10^{33}$~erg~s$^{-1}$, assuming  a distance of 6~kpc. 

Interestingly, the PWN in G292.0+1.8 shows a high luminosity in the radio band.  
Its radio-to-TeV luminosity ratio, $L_r= L_{\text{radio}} / L_\gamma(\text{TeV}) \geq 0.4$, is among the highest in the known population, placing it alongside  3C~58 and the Crab Nebula. 
Similarly, the  X-ray-to-TeV luminosity ratio, $L_{\mathrm{X}}= L_{\text{X-ray}} / L_\gamma(\text{TeV}) 
\geq 40$, is comparable to that of 3C~58 \citep{an2019}. 
Only three other PWNe exhibit a larger  $L_{\mathrm{X}}$ ratio: G21.5$-$0.9 and the Crab Nebula in the Galaxy, and N~158A in the Large Magellanic Cloud \citep{zhu2018}.   
These high $L_r$ and $L_{\mathrm{X}}$ values suggest that the nebula is an efficient emitter in the radio and X-ray bands, but a relatively weak source at TeV energies. 
Indeed, the nebula associated with PSR~J1124$-$5916 in G292.0+1.8 appears as an outlier in the established correlations between spin-down power and both radio and X-ray luminosities. It lies well outside the 2$\sigma$ confidence bands of the $L_r$-$\dot{E}$ and $L_{\mathrm{X}}$-$\dot{E}$ correlations defined for Galactic PWNe 
powered by high spin-down luminosity pulsars, most of which are also effective TeV emitters \citep{zhu2018}.  
Nevertheless, the detection of significant GeV emission from the PWN in G292.0+1.8 indicates that its particles can radiate efficiently at high energies (see Sect.~\ref{sec:3GeVemission}). This behaviour offers valuable insight into the energy-dependent properties of particle acceleration and cooling in young PWNe.
\begin{figure}[h!]
    \includegraphics[scale=0.6]{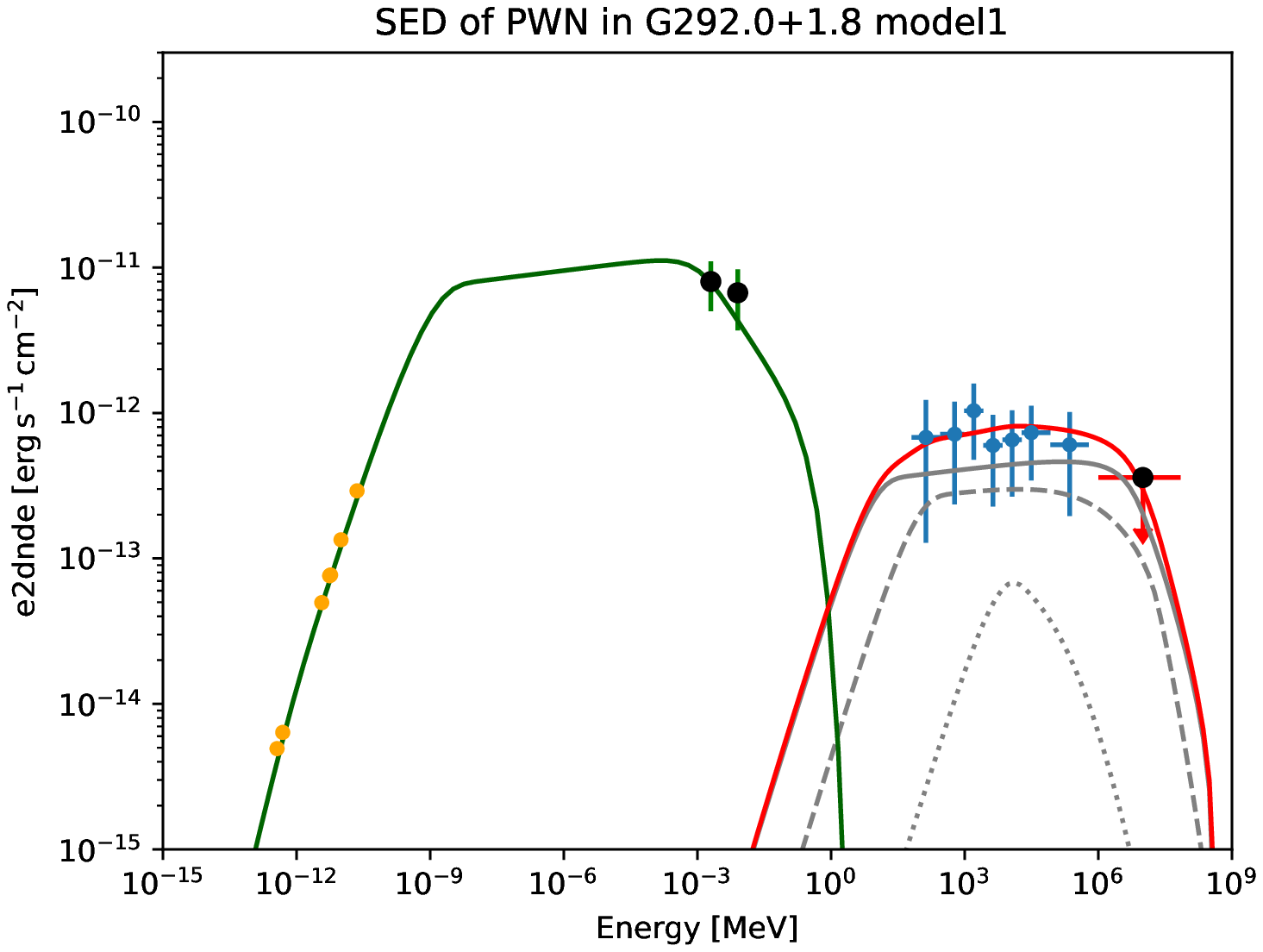}
    \includegraphics[scale=0.6]{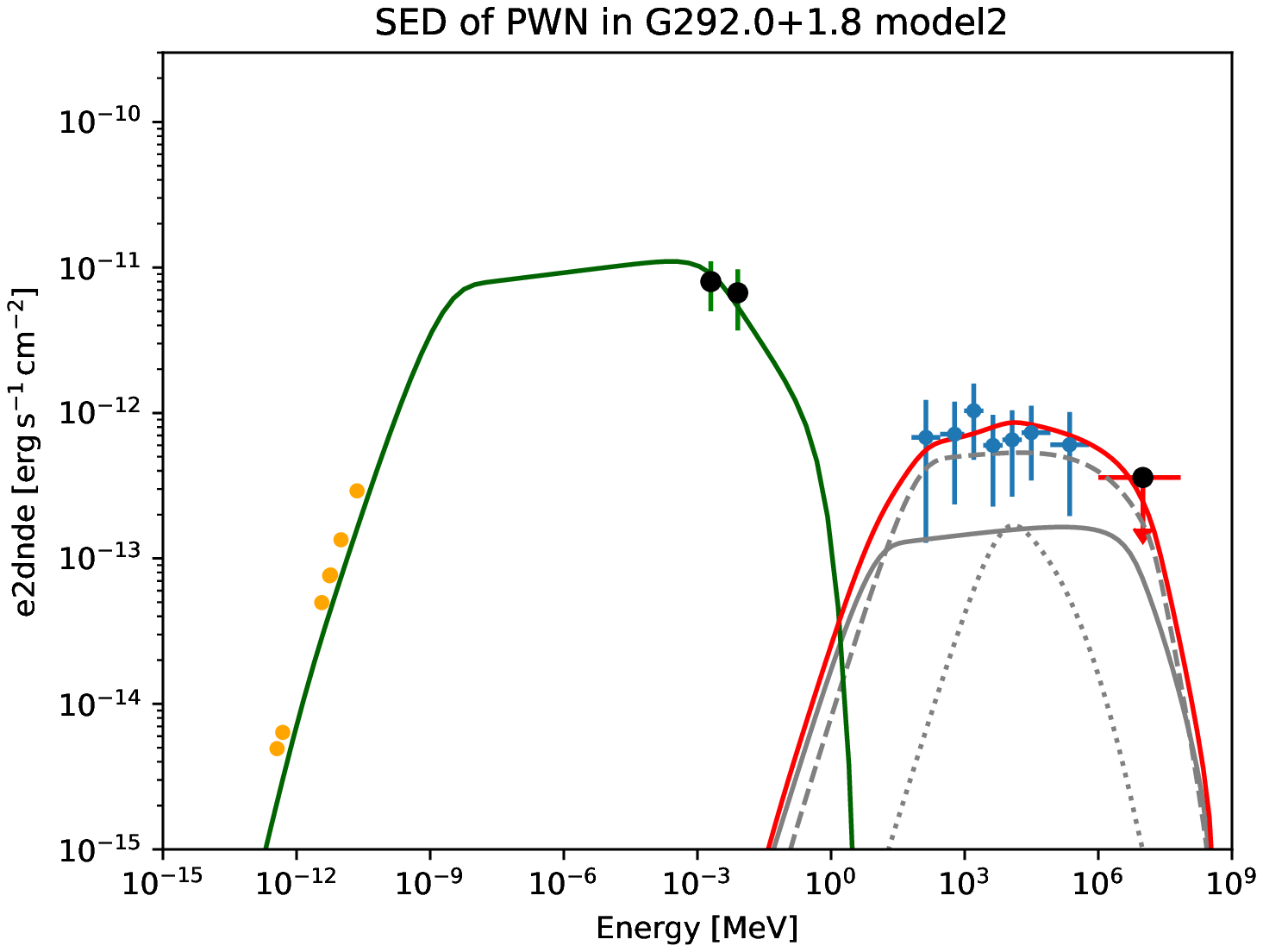}
 \caption{SED of the G292.0+1.8 PWN emission, fitted with the one-zone electron model described in 
 Sect.~\ref{sec:4discussion}. 
 The dataset includes radio emission derived from \citet{gaensler2003} and Sect.~\ref{sec:2radio-data}, the total PWN X-ray spectrum from \citep{hughes2001}, 
 the VHE upper limit derived by \citep{2018_HESS_PWN_Pop}, and the \it Fermi\rm-LAT spectrum obtained from our analysis. 
 Two models, Model 1 and Model 2, are used to fit the data, with their parameters summarised in Table~\ref{tab:spectra_Pulsar_table}.  The IC emission (black line) is computed taking into account contributions from the CMB (dashed line), starlight (dotted line), and dust (dotted-dashed line) targets, following the energy density estimates from \citet{porter2006} and \citet{Shibata+11} for Model 1, and \citet{zhu2018} for Model 2.
}
 \label{fig:SED_Spectra}
\end{figure}

When comparing PWNe powered by pulsars with similar ages and spin-down luminosities, it becomes evident that their broadband emission properties can differ significantly. For instance, the PWN in G292.0+1.8 resembles 3C~58 in several key respects: 
both are associated with energetic, young pulsars -- $\sim$2,500--2,900~yr old in the case of  G292.0+1.8 \citep{Camilo2002G292,temim2022} and between $\sim$830--2,500~yr \citep{fesen2008,kothes2013} and both nebulae are characterised by high radio and X-ray luminosities \citep{zhu2018}.  However, a striking difference emerges at $\gamma$-ray energies: while both nebulae are detected in the GeV range, only 3C~58 has been observed at TeV energies. One key factor that may contribute to this divergence is the interaction (or lack thereof) with the reverse shock of the host SNR. G292.0+1.8 is thought to have already formed a reverse shock, although this interpretation remains under debate.  
\citet{gaensler2003} suggested that the reverse shock has reached the PWN, based on the apparent lack of a clear spatial separation between the nebula and the SNR shell. Conversely, \citet{bhalerao2015} argued that much of the ejecta have not yet been impacted by the reverse shock, implying that the PWN may still be in a pre-interaction stage. 
This uncertainty complicates the interpretation of the PWN’s broadband emission and introduces challenges in modelling its evolution. In the case of 3C~58, it also remains unclear whether the PWN has been compressed by a reverse shock propagating back towards the pulsar. The question of whether such an interaction has occurred is still open, and resolving it is crucial for understanding the current state and long-term evolution of the nebula. 
The PWN in G292.0+1.8 also shares characteristics with that in SNR~G54.1+0.3, particularly regarding the properties of their pulsars. Both pulsars have similar estimated ages (2,000--2,900~yr for G54.1+0.3, \citealt{Camilo2002G292,temim2022,gelfand2015, Camilo2002G54}) and spin-down powers  ($\dot{E} = 1.2 \times 10^{37}~\mathrm{erg}\,\mathrm{s}^{-1}$ \citealt{Camilo2002G292,Camilo2002G54}). In addition, both systems lie at similar distances of  around 6~kpc \citep{Long2022G292, Leahy2008G54}. Despite these similarities, however, G54.1+0.3 remains undetected in the GeV band, although its TeV emission is relatively strong.    
 The discrepancies observed in the broadband SEDs of these three systems 
 --even though they are powered by pulsars with similar characteristics-- 
 suggest that $\gamma$-ray emission in young PWNe is shaped not solely by intrinsic pulsar properties, but also by environmental factors. These may include the density and spectrum of local photon fields, the strength and geometry of the magnetic field, and ambient  interstellar conditions within the host SNR. 
 Such diversity can impact particle cooling, diffusion, and the efficiency of IC scattering, ultimately affecting whether a given PWN appears GeV-bright, TeV-bright, or both.

\subsection{Broadband SED}
\label{subsec:broadbandSED}
Figure~\ref{fig:SED_Spectra} shows the broadband SED of the PWN in G292.0+1.8, incorporating the radio data from this work (Sect.~\ref{sec:2radio-data}), X-ray measurements from \it Chandra \rm \citep{hughes2001}, LAT  fluxes from our own analysis, and the TeV upper limit derived from HESS observations \citep{2018_HESS_PWN_Pop}. 
LAT flux uncertainties include both statistical and systematic contributions. 
A compact torus-like structure surrounding PSR~J1124$-$5916 has been detected in the optical and infrared  \citep{zharikov2008,zyuzin2009}; however, this feature is considerably smaller than the full extent of the radio and X-ray nebula and does not dominate the total PWN emission. For this reason, optical and IR data are not included in the SED.

To account for the broadband emission, we modelled the SED of G292.0+1.8 as a particle-dominated nebula powered by a single electron population. 
We employed the \it Naima \rm package \citep{Zabalza+15} to reproduce the nonthermal photon spectrum, assuming synchrotron radiation in a homogeneous and constant magnetic field, as well as  IC scattering of electrons interacting with ambient photon fields.
These target fields include the Cosmic Microwave Background (CMB), stellar light (characterised by an IR component at 26.5~K), and dust re-emission in the NIR (at 2,800~K).

Due to G292.0+1.8's large Galactocentric distance (R$\sim$14.8~kpc), the  local interstellar radiation field (ISRF) at $z$= 0 is expected to be relatively weak.
According to  \it GALPROP \rm simulations, typical ISRF values at this location are U$_{\mathrm{FIR}}=0.2$~eV~cm$^{-3}$ for the far-infrared (FIR) component and  U$_{\mathrm{NIR}}=0.1$~eV~cm$^{-3}$ for the NIR component \citep{strong2000,Shibata+11}. However, several studies adopt enhanced ISRFs to fit PWN spectra, such as U$_{\mathrm{FIR}}=0.42$~eV~cm$^{-3}$ and U$_\mathrm{NIR}=0.7$~eV~cm$^{-3}$ \citep{tanaka2013}, or even U$_{\mathrm{FIR}}=1.0$~eV~cm$^{-3}$ and U$_{\mathrm{NIR}}=0.7$~eV~cm$^{-3}$ \citep{zhu2018}. In our modelling, we considered two scenarios. 
Model 1 assumes a standard ISRF configuration with U$_{\mathrm{FIR}}=0.2$~eV~cm$^{-3}$ and U$_{\mathrm{NIR}}=0.1$~eV~cm$^{-3}$, consistent with \it GALPROP \rm predictions. 
Model 2 adopts an enhanced ISRF, with U$_{\mathrm{FIR}}=1.0$~eV~cm$^{-3}$ and U$_{\mathrm{NIR}}=0.7$~eV~cm$^{-3}$, to test the impact of elevated ambient photon densities on the IC component of the emission.

The synchrotron component of the SED  is strongly constrained by the hard spectral index observed in the radio band. For a synchrotron radio spectrum with an index $\alpha \simeq -0.012$, the corresponding  
power-law index of the injected electron spectrum index is  
$\gamma_1= 1 - 2 \alpha   \simeq 1.024$ (with $\gamma_1 > 0$ ). 
A comparison between the radio and X-ray bands indicates a spectral break that cannot be attributed solely to synchrotron cooling. This necessitates the introduction of an intrinsic cut-off in the electron injection spectrum at around $E_{\mathrm{injec, elec}} \simeq40$~GeV, beyond which the electron distribution steepens to $\gamma_2 \sim 2.9$. This behaviour is similar to that observed in the  PWN 3C~58 \citep{slane2008}, as well as in other modelling studies, such as those  by \citet{zhu2018} and \citet{liu2024}. The spectral break consistently occurs  between the radio and X-ray bands, lying close to the infrared regime.

Additionally, the electron spectrum features a second break at $E_{\mathrm{b_{cool},e}} \simeq 25$~TeV, resulting from synchrotron losses within the nebula. This high-energy cut-off is well constrained, appearing just below the X-ray band in the synchrotron emission, and between the LAT spectrum and the very-high-energy (VHE) upper limits in the IC emission. The overall shape of the  electron spectrum needed to reproduce the SED is a double broken power-law  characterised by two energy breaks and three spectral indices (summarised in Table~\ref{tab:parameters}): 
\[
N(E) =
\begin{cases}
N_0 \cdot E^{-\gamma_1} & \text{si } E < E_{\mathrm{b_{inj},e}} \\
N_0 \cdot E_{\mathrm{b_{inj},e}}^{\gamma_2 - \gamma_1} \cdot E^{-\gamma_2} & \text{si } E_{\mathrm{b_{inj},e}} \leq E < E_{\mathrm{b_{cool},e}} \\
N_0 \cdot E_{\mathrm{b_{inj},e}}^{\gamma_2 - \gamma_1} \cdot E_{\mathrm{b_{cool},e}}^{\gamma_{\mathrm{cool}} - \gamma_2} \cdot E^{-\gamma_{\mathrm{cool}}} & \text{si } E \geq E_{\mathrm{b_{cool},e}}
\end{cases}
\]
Notably, 
the relation $\gamma_{\mathrm{cool}}= \gamma_2 + 1 $ as well as the location of the cooling energy break are in agreement with a synchrotron-loss-dominated PWN in a homogeneous and constant magnetic field.  
In this case, the synchrotron cooling break energy (in eV), for a PWN of age $t_{\text{age}}$ (in yr) and magnetic field strength $B$ (in G), can be approximated by the relation  $E_{\text{cool,synch}}(\mathrm{eV}) = 1.7 \times 10^9 \, B^{-3} \, t_\text{age}^{-2}$ \citep{slane2008}. For a field strength of $B \simeq 15$~$\mu$G and a typical age of $t_{\mathrm{age}}\sim$ 2,5~kyr, the synchrotron break lies around 0.4~keV, corresponding to an electron break energy of $E_{\mathrm{b_{cool},e}} \sim20$~TeV. This estimate is rather compatible with our parameters in Model 1 and consistent with values derived in other models, such as \citet{liu2024}. However it appears difficult to reach the order of magnitude of the energy cooling break with the higher magnetic field strength used in Model 2. 

The magnetic field strength of about 15~$\mu$G in Model 1 appears to depart from equipartition, yet it remains compatible with the age and size of the PWN, and falls within the range of values  derived in previous PWN models. For instance, \citet{tanaka2013} estimated a field of  $B = 16$~$\mu$G, while \citet{martin2014} found a slightly higher value of $B = 21$~$\mu$G, the latter being closer to our Model 2. Comparable magnetic field values have also been inferred for the  PWN 3C~58 \citep{slane2008,tanaka2013}, reinforcing the plausibility of our estimates. 

The total energy contained in the electron population is estimated to be $W_{\text{e}}(\,>2\, \text{GeV}) = 5.3 \times 10^{48}$~erg for Model 1 and $W_{\text{e}}(\, >2\, \text{GeV}) = 2.2 \times 10^{48}$~erg for Model 2. The very high rotational energy of PSR~J1124$-$5916 is sufficient to account for the energetics of the nebula over its 2,500-yr lifetime. 

Our modelling indicates that the emission can be consistently explained by a single electron population characterised by two spectral breaks. The first, occurring at a few tens of GeV, is likely intrinsic to the particle injection mechanism, while the second arises from synchrotron cooling in a magnetic field of $\sim15$~$\mu$G. This field strength, although far from equipartition, is similar to that inferred for the 3C~58 nebula. Given these similarities, including the spectral morphology and inferred magnetic conditions, we favour a scenario in which the reverse shock has not yet impacted the PWN in G292.0+1.8, analogous to the evolutionary state proposed for 3C~58.
\begin{table}[h!]
\caption{Models parameters.
} 
\small\centering
%-----------------------------------------------------------------
\begin{tabular}{lcc}
\hline\hline
%------------------------------------------------------------------
  Parameters         &  Values    \\
   \hline   \hline
\multicolumn{2}{c}{Electrons spectrum} \\ 
    \hline
    E$_{\mathrm{min}}$  &	 1~GeV  \\ 
    E$_{\mathrm{max}}$ &   500~TeV  \\
    $\gamma_1$ &   1.024 \\
    $\gamma_2$ &   $\sim$ 2.9  \\
    $\gamma_{\mathrm{cool}}$  &   $\sim$ 3.9  \\
    E$_{\mathrm{b_{inj},e}}$ &    $\sim$ 40~GeV \\
    E$_{\mathrm{b_{cool},e}}$ &   $\sim$ 25~TeV  \\
    \hline 
    \multicolumn{2}{c}{Model 1} \\
    \hline
    FIR energy density &     0.2~eV~cm$^{-3}$ \\ 
    NIR energy density &     0.1~eV~cm$^{-3}$ \\  
    Magnetic field & $\sim 15$~$\mu$G \\ 
    We( > 1~GeV) &   $\sim 5.3 \times 10^{48}$~erg \\
    \hline 
    \multicolumn{2}{c}{Model 2} \\
    \hline
    FIR energy density &     1.0~eV~cm$^{-3}$\\ 
    NIR energy density &     0.7~eV~cm$^{-3}$\\  
    %\hline
    Magnetic field & $\sim 25$~$\mu$G \\ 
    We( > 1~GeV) &    $\sim 2.2 \times 10^{48}$~erg \\
    \hline    
 \end{tabular}
 \label{tab:parameters}
 \tablefoot{Detail of models parameters presented in the SED Fig.~\ref{fig:SED_Spectra}. Model 1 and 2 use the same electron spectrum shape.}
\end{table}

\section{Summary}
\label{sec:5conclusions}
The limited number of PWNe firmly detected by \textit{Fermi}-LAT underscores the pressing for new identifications to better constrain their SEDs in the GeV domain. This task is particularly challenging due to the overwhelming brightness of the associated pulsars, whose magnetospheric emission often outshines the comparatively fainter nebular component--potentially biasing  the pulsar spectrum if nebular contamination is not properly accounted for.
As demonstrated in this work, time-resolved (phase-resolved) 
$\gamma$-ray analysis is a powerful diagnostic tool for disentangling these components, even in cases where spatial separation is not feasible with the current  angular resolution of  $\gamma$-ray instruments.

In this study, we achieved several key results:

\begin{enumerate}[leftmargin=1.5em]
\item We reported the first detection of GeV $\gamma$-ray emission potentially associated with the PWN in G292.0+1.8, successfully disentangling it from the bright pulsed emission of the nearby \textit{Fermi}-LAT pulsar, 4FGL~J1124$-$5916. This places G292.0+1.8 among the limited number of young pulsar–PWN systems for which the unpulsed nebular component has been spectrally isolated from the magnetospheric signal in the GeV regime.

\item We characterised the pulsed $\gamma$-ray spectra of the two main peaks (P1 and P2) of PSR~J1124$-$5916, refining the understanding of its magnetospheric emission.

\item We extracted and measure the spectrum of the unpulsed $\gamma$-ray component in the GeV band.

\item We constructed and modelled the broadband SED of the PWN candidate,  by combining the newly extracted 
$\gamma$-ray spectrum with radio and X-ray data. 
The emission is well described by a single electron population with two spectral breaks:  
one intrinsic to the injection spectrum around a few tens of GeV, and another due to synchrotron cooling in a magnetic field of $\sim$15\,$\mu$G. 
This relatively low magnectic field, significantly below equipartition, is reminiscent of the value inferred for  3C~58, supporting the notion that some young PWNe evolve in low-field environments.
\end{enumerate}

The PWN in G292.0+1.8 exhibits a notably low TeV flux relative to its spin-down power, a feature it shares with 3C~58 and that contrasts with other young systems such as G54.1+0.3, which shows much stronger TeV emission.  A precise characterisation of the TeV spectrum will be crucial for constraining the high-energy cutoff of the electron distribution and verifying the synchrotron cooling break inferred from our model.
More broadly, the identification of GeV emission from the unpulsed component, which our \textit{Fermi}-LAT analysis suggests is associated with the PWN in G292.0+1.8, adds to the  still limited sample of young systems where such separation is achievable. This result provides valuable insights into the spectral diversity and evolutionary behaviour of pulsars embedded in distinct environments. The ability to isolate and characterise both the pulsed and nebular components in a single PSR–PWN–SNR system like G292.0+1.8 represents a meaningful advance for multiwavelength studies aimed at resolving the interplay between compact objects and their surrounding media.

\begin{acknowledgements} 
This work and collaboration is supported by the International Emerging Actions program from CNRS (France). 
GC and NM are members of the Carrera del Investigador Científico (CONICET, Argentina). CONICET has partially supported this research (PIP 11220220100337). 
This paper used archived data obtained through the Galactic and Extragalactic All-sky Murchison Widefield Array, the CSIRO ASKAP Science Data Archive (CASDA), and the SARAO MeerKAT 1.3~GHz Galactic Plane Survey.
The authors thank Maxime Regeard and Arache Djannati-Atai for fruitful discussions regarding pulsar observations with \textit{Fermi}-LAT, which helped improve the quality of this work.
\end{acknowledgements}

\bibliographystyle{aa}
 \bibliography{biblio-3SNR.bib}

\begin{thebibliography}{60}
\expandafter\ifx\csname natexlab\endcsname\relax\def\natexlab#1{#1}\fi

\bibitem[{{Abdo} {et~al.}(2010){Abdo}, {Ackermann}, {Ajello}, {Atwood}, {Axelsson}, {Baldini}, {Ballet}, {Barbiellini}, {Baring}, {Bastieri}, {Baughman}, {Bechtol}, {Bellazzini}, {Berenji}, {Blandford}, {Bloom}, {Bonamente}, {Borgland}, {Bregeon}, {Brez}, {Brigida}, {Bruel}, {Burnett}, {Buson}, {Caliandro}, {Cameron}, {Camilo}, {Caraveo}, {Casandjian}, {Cecchi}, {{\c{C}}elik}, {Charles}, {Chekhtman}, {Cheung}, {Chiang}, {Ciprini}, {Claus}, {Cognard}, {Cohen-Tanugi}, {Cominsky}, {Conrad}, {Corbet}, {Cutini}, {den Hartog}, {Dermer}, {de Angelis}, {de Luca}, {de Palma}, {Digel}, {Dormody}, {Silva}, {Drell}, {Dubois}, {Dumora}, {Espinoza}, {Farnier}, {Favuzzi}, {Fegan}, {Ferrara}, {Focke}, {Fortin}, {Frailis}, {Freire}, {Fukazawa}, {Funk}, {Fusco}, {Gargano}, {Gasparrini}, {Gehrels}, {Germani}, {Giavitto}, {Giebels}, {Giglietto}, {Giommi}, {Giordano}, {Glanzman}, {Godfrey}, {Gotthelf}, {Grenier}, {Grondin}, {Grove}, {Guillemot}, {Guiriec}, {Gwon}, {Hanabata}, {Harding}, {Hayashida}, {Hays}, {Hughes}, {Jackson},
  {J{\'o}hannesson}, {Johnson}, {Johnson}, {Johnson}, {Johnson}, {Johnston}, {Kamae}, {Kanbach}, {Kaspi}, {Katagiri}, {Kataoka}, {Kawai}, {Kerr}, {Kn{\"o}dlseder}, {Kocian}, {Kramer}, {Kuss}, {Lande}, {Latronico}, {Lemoine-Goumard}, {Livingstone}, {Longo}, {Loparco}, {Lott}, {Lovellette}, {Lubrano}, {Lyne}, {Madejski}, {Makeev}, {Manchester}, {Marelli}, {Mazziotta}, {McConville}, {McEnery}, {McGlynn}, {Meurer}, {Michelson}, {Mineo}, {Mitthumsiri}, {Mizuno}, {Moiseev}, {Monte}, {Monzani}, {Morselli}, {Moskalenko}, {Murgia}, {Nakamori}, {Nolan}, {Norris}, {Noutsos}, {Nuss}, {Ohsugi}, {Omodei}, {Orlando}, {Ormes}, {Ozaki}, {Paneque}, {Panetta}, {Parent}, {Pelassa}, {Pepe}, {Pesce-Rollins}, {Piron}, {Porter}, {Rain{\`o}}, {Rando}, {Ransom}, {Ray}, {Razzano}, {Rea}, {Reimer}, {Reimer}, {Reposeur}, {Ritz}, {Rodriguez}, {Romani}, {Roth}, {Ryde}, {Sadrozinski}, {Sanchez}, {Sander}, {Saz Parkinson}, {Scargle}, {Schalk}, {Sellerholm}, {Sgr{\`o}}, {Siskind}, {Smith}, {Smith}, {Spandre}, {Spinelli}, {Stappers}, {Starck},
  {Striani}, {Strickman}, {Strong}, {Suson}, {Tajima}, {Takahashi}, {Takahashi}, {Tanaka}, {Thayer}, {Thayer}, {Theureau}, {Thompson}, {Thorsett}, {Tibaldo}, {Tibolla}, {Torres}, {Tosti}, {Tramacere}, {Uchiyama}, {Usher}, {Van Etten}, {Vasileiou}, {Venter}, {Vilchez}, {Vitale}, {Waite}, {Wang}, {Wang}, {Watters}, {Weltevrede}, {Winer}, {Wood}, {Ylinen}, \& {Ziegler}}]{Abdo2010}
{Abdo}, A.~A., {Ackermann}, M., {Ajello}, M., {et~al.} 2010, \apjs, 187, 460

\bibitem[{Abdo {et~al.}(2009)Abdo, Ackermann, Ajello, Bechtol, Berenji, Blandford, Bloom, Borgland, Bouvier, Baldini, Bellazzini, Bregeon, Brez, Ballet, Barbiellini, Baring, Bastieri, Baughman, Bonamente, \& Brigida}]{Abdo2009_W51C}
Abdo, A.~A., Ackermann, M., Ajello, M., {et~al.} 2009, \apj, 706, L1

\bibitem[{Abdo {et~al.}(2013)Abdo, Ajello, Allafort, Baldini, Ballet, Barbiellini, Baring, Bastieri, Belfiore, Bellazzini, Bhattacharyya, Bissaldi, Bloom, Bonamente, Bottacini, Brandt, Bregeon, Brigida, Bruel, Buehler, Burgay, Burnett, Busetto, Buson, Caliandro, Cameron, Camilo, Caraveo, Casandjian, Cecchi, Ãelik, Charles, Chaty, Chaves, Chekhtman, Chen, Chiang, Chiaro, Ciprini, Claus, Cognard, Cohen-Tanugi, Cominsky, Conrad, Cutini, D'Ammando, de~Angelis, DeCesar, Luca, den Hartog, de~Palma, Dermer, Desvignes, Digel, Venere, Drell, Drlica-Wagner, Dubois, Dumora, Espinoza, Falletti, Favuzzi, Ferrara, Focke, Franckowiak, Freire, Funk, Fusco, Gargano, Gasparrini, Germani, Giglietto, Giommi, Giordano, Giroletti, Glanzman, Godfrey, Gotthelf, Grenier, Grondin, Grove, Guillemot, Guiriec, Hadasch, Hanabata, Harding, Hayashida, Hays, Hessels, Hewitt, Hill, Horan, Hou, Hughes, Jackson, Janssen, Jogler, JÃ³hannesson, Johnson, Johnson, Johnson, Johnson, Johnston, Kamae, Kataoka, Keith, Kerr, KnÃ¶dlseder, Kramer,
  Kuss, Lande, Larsson, Latronico, Lemoine-Goumard, Longo, Loparco, Lovellette, Lubrano, Lyne, Manchester, Marelli, Massaro, Mayer, Mazziotta, McEnery, McLaughlin, Mehault, Michelson, Mignani, Mitthumsiri, Mizuno, Moiseev, Monzani, Morselli, Moskalenko, Murgia, Nakamori, Nemmen, Nuss, Ohno, Ohsugi, Orienti, Orlando, Ormes, Paneque, Panetta, Parent, Perkins, Pesce-Rollins, Pierbattista, Piron, Pivato, Pletsch, Porter, Possenti, RainÃ², Rando, Ransom, Ray, Razzano, Rea, Reimer, Reimer, Renault, Reposeur, Ritz, Romani, Roth, Rousseau, Roy, Ruan, Sartori, Parkinson, Scargle, Schulz, SgrÃ², Shannon, Siskind, Smith, Spandre, Spinelli, Stappers, Strong, Suson, Takahashi, Thayer, Thayer, Theureau, Thompson, Thorsett, Tibaldo, Tibolla, Tinivella, Torres, Tosti, Troja, Uchiyama, Usher, Vandenbroucke, Vasileiou, Venter, Vianello, Vitale, Wang, Weltevrede, Winer, Wolff, Wood, Wood, Wood, \& Yang}]{Abdo_2013}
Abdo, A.~A., Ajello, M., Allafort, A., {et~al.} 2013, The Astrophysical Journal Supplement Series, 208, 17

\bibitem[{{Abdollahi} {et~al.}(2022{\natexlab{a}}){Abdollahi}, {Acero}, {Baldini}, {Ballet}, {Bastieri}, {Bellazzini}, {Berenji}, {Berretta}, {Bissaldi}, {Blandford}, {Bloom}, {Bonino}, {Brill}, {Britto}, {Bruel}, {Burnett}, {Buson}, {Cameron}, {Caputo}, {Caraveo}, {Castro}, {Chaty}, {Cheung}, {Chiaro}, {Cibrario}, {Ciprini}, {Coronado-Bl{\'a}zquez}, {Crnogorcevic}, {Cutini}, {D'Ammando}, {De Gaetano}, {Digel}, {Di Lalla}, {Dirirsa}, {Di Venere}, {Dom{\'\i}nguez}, {Fallah Ramazani}, {Fegan}, {Ferrara}, {Fiori}, {Fleischhack}, {Franckowiak}, {Fukazawa}, {Funk}, {Fusco}, {Galanti}, {Gammaldi}, {Gargano}, {Garrappa}, {Gasparrini}, {Giacchino}, {Giglietto}, {Giordano}, {Giroletti}, {Glanzman}, {Green}, {Grenier}, {Grondin}, {Guillemot}, {Guiriec}, {Gustafsson}, {Harding}, {Hays}, {Hewitt}, {Horan}, {Hou}, {J{\'o}hannesson}, {Karwin}, {Kayanoki}, {Kerr}, {Kuss}, {Landriu}, {Larsson}, {Latronico}, {Lemoine-Goumard}, {Li}, {Liodakis}, {Longo}, {Loparco}, {Lott}, {Lubrano}, {Maldera}, {Malyshev}, {Manfreda},
  {Mart{\'\i}-Devesa}, {Mazziotta}, {Mereu}, {Meyer}, {Michelson}, {Mirabal}, {Mitthumsiri}, {Mizuno}, {Moiseev}, {Monzani}, {Morselli}, {Moskalenko}, {Negro}, {Nuss}, {Omodei}, {Orienti}, {Orlando}, {Paneque}, {Pei}, {Perkins}, {Persic}, {Pesce-Rollins}, {Petrosian}, {Pillera}, {Poon}, {Porter}, {Principe}, {Rain{\`o}}, {Rando}, {Rani}, {Razzano}, {Razzaque}, {Reimer}, {Reimer}, {Reposeur}, {S{\'a}nchez-Conde}, {Saz Parkinson}, {Scotton}, {Serini}, {Sgr{\`o}}, {Siskind}, {Smith}, {Spandre}, {Spinelli}, {Sueoka}, {Suson}, {Tajima}, {Tak}, {Thayer}, {Thompson}, {Torres}, {Troja}, {Valverde}, {Wood}, \& {Zaharijas}}]{abbollahi2022}
{Abdollahi}, S., {Acero}, F., {Baldini}, L., {et~al.} 2022{\natexlab{a}}, \apjs, 260, 53

\bibitem[{{Abdollahi} {et~al.}(2022{\natexlab{b}}){Abdollahi}, {Acero}, {Baldini}, {Ballet}, {Bastieri}, {Bellazzini}, {Berenji}, {Berretta}, {Bissaldi}, {Blandford}, {Bloom}, {Bonino}, {Brill}, {Britto}, {Bruel}, {Burnett}, {Buson}, {Cameron}, {Caputo}, {Caraveo}, {Castro}, {Chaty}, {Cheung}, {Chiaro}, {Cibrario}, {Ciprini}, {Coronado-Bl{\'a}zquez}, {Crnogorcevic}, {Cutini}, {D'Ammando}, {De Gaetano}, {Digel}, {Di Lalla}, {Dirirsa}, {Di Venere}, {Dom{\'\i}nguez}, {Fallah Ramazani}, {Fegan}, {Ferrara}, {Fiori}, {Fleischhack}, {Franckowiak}, {Fukazawa}, {Funk}, {Fusco}, {Galanti}, {Gammaldi}, {Gargano}, {Garrappa}, {Gasparrini}, {Giacchino}, {Giglietto}, {Giordano}, {Giroletti}, {Glanzman}, {Green}, {Grenier}, {Grondin}, {Guillemot}, {Guiriec}, {Gustafsson}, {Harding}, {Hays}, {Hewitt}, {Horan}, {Hou}, {J{\'o}hannesson}, {Karwin}, {Kayanoki}, {Kerr}, {Kuss}, {Landriu}, {Larsson}, {Latronico}, {Lemoine-Goumard}, {Li}, {Liodakis}, {Longo}, {Loparco}, {Lott}, {Lubrano}, {Maldera}, {Malyshev}, {Manfreda},
  {Mart{\'\i}-Devesa}, {Mazziotta}, {Mereu}, {Meyer}, {Michelson}, {Mirabal}, {Mitthumsiri}, {Mizuno}, {Moiseev}, {Monzani}, {Morselli}, {Moskalenko}, {Negro}, {Nuss}, {Omodei}, {Orienti}, {Orlando}, {Paneque}, {Pei}, {Perkins}, {Persic}, {Pesce-Rollins}, {Petrosian}, {Pillera}, {Poon}, {Porter}, {Principe}, {Rain{\`o}}, {Rando}, {Rani}, {Razzano}, {Razzaque}, {Reimer}, {Reimer}, {Reposeur}, {S{\'a}nchez-Conde}, {Saz Parkinson}, {Scotton}, {Serini}, {Sgr{\`o}}, {Siskind}, {Smith}, {Spandre}, {Spinelli}, {Sueoka}, {Suson}, {Tajima}, {Tak}, {Thayer}, {Thompson}, {Torres}, {Troja}, {Valverde}, {Wood}, \& {Zaharijas}}]{4FGLDR32022}
{Abdollahi}, S., {Acero}, F., {Baldini}, L., {et~al.} 2022{\natexlab{b}}, \apjs, 260, 53

\bibitem[{{Acero} {et~al.}(2013){Acero}, {Ackermann}, {Ajello}, {Allafort}, {Baldini}, {Ballet}, {Barbiellini}, {Bastieri}, {Bechtol}, {Bellazzini}, {Blandford}, {Bloom}, {Bonamente}, {Bottacini}, {Brandt}, {Bregeon}, {Brigida}, {Bruel}, {Buehler}, {Buson}, {Caliandro}, {Cameron}, {Caraveo}, {Cecchi}, {Charles}, {Chaves}, {Chekhtman}, {Chiang}, {Chiaro}, {Ciprini}, {Claus}, {Cohen-Tanugi}, {Conrad}, {Cutini}, {Dalton}, {D'Ammando}, {de Palma}, {Dermer}, {Di Venere}, {Silva}, {Drell}, {Drlica-Wagner}, {Falletti}, {Favuzzi}, {Fegan}, {Ferrara}, {Focke}, {Franckowiak}, {Fukazawa}, {Funk}, {Fusco}, {Gargano}, {Gasparrini}, {Giglietto}, {Giordano}, {Giroletti}, {Glanzman}, {Godfrey}, {Gr{\'e}goire}, {Grenier}, {Grondin}, {Grove}, {Guiriec}, {Hadasch}, {Hanabata}, {Harding}, {Hayashida}, {Hayashi}, {Hays}, {Hewitt}, {Hill}, {Horan}, {Hou}, {Hughes}, {Inoue}, {Jackson}, {Jogler}, {J{\'o}hannesson}, {Johnson}, {Kamae}, {Kawano}, {Kerr}, {Kn{\"o}dlseder}, {Kuss}, {Lande}, {Larsson}, {Latronico}, {Lemoine-Goumard},
  {Longo}, {Loparco}, {Lovellette}, {Lubrano}, {Marelli}, {Massaro}, {Mayer}, {Mazziotta}, {McEnery}, {Mehault}, {Michelson}, {Mitthumsiri}, {Mizuno}, {Monte}, {Monzani}, {Morselli}, {Moskalenko}, {Murgia}, {Nakamori}, {Nemmen}, {Nuss}, {Ohsugi}, {Okumura}, {Orienti}, {Orlando}, {Ormes}, {Paneque}, {Panetta}, {Perkins}, {Pesce-Rollins}, {Piron}, {Pivato}, {Porter}, {Rain{\`o}}, {Rando}, {Razzano}, {Reimer}, {Reimer}, {Reposeur}, {Ritz}, {Roth}, {Rousseau}, {Saz Parkinson}, {Schulz}, {Sgr{\`o}}, {Siskind}, {Smith}, {Spandre}, {Spinelli}, {Suson}, {Takahashi}, {Takeuchi}, {Thayer}, {Thayer}, {Thompson}, {Tibaldo}, {Tibolla}, {Tinivella}, {Torres}, {Tosti}, {Troja}, {Uchiyama}, {Vandenbroucke}, {Vasileiou}, {Vianello}, {Vitale}, {Werner}, {Winer}, {Wood}, \& {Yang}}]{FERMI_PWNTeV2013}
{Acero}, F., {Ackermann}, M., {Ajello}, M., {et~al.} 2013, \apj, 773, 77

\bibitem[{{Ackermann} {et~al.}(2012{\natexlab{a}}){Ackermann}, {Ajello}, {Albert}, {Allafort}, {Atwood}, {Axelsson}, {Baldini}, {Ballet}, {Barbiellini}, {Bastieri}, {Bechtol}, {Bellazzini}, {Bissaldi}, {Blandford}, {Bloom}, {Bogart}, {Bonamente}, {Borgland}, {Bottacini}, {Bouvier}, {Brandt}, {Bregeon}, {Brigida}, {Bruel}, {Buehler}, {Burnett}, {Buson}, {Caliandro}, {Cameron}, {Caraveo}, {Casandjian}, {Cavazzuti}, {Cecchi}, {{\c{C}}elik}, {Charles}, {Chaves}, {Chekhtman}, {Cheung}, {Chiang}, {Ciprini}, {Claus}, {Cohen-Tanugi}, {Conrad}, {Corbet}, {Cutini}, {D'Ammando}, {Davis}, {de Angelis}, {DeKlotz}, {de Palma}, {Dermer}, {Digel}, {Silva}, {Drell}, {Drlica-Wagner}, {Dubois}, {Favuzzi}, {Fegan}, {Ferrara}, {Focke}, {Fortin}, {Fukazawa}, {Funk}, {Fusco}, {Gargano}, {Gasparrini}, {Gehrels}, {Giebels}, {Giglietto}, {Giordano}, {Giroletti}, {Glanzman}, {Godfrey}, {Grenier}, {Grove}, {Guiriec}, {Hadasch}, {Hayashida}, {Hays}, {Horan}, {Hou}, {Hughes}, {Jackson}, {Jogler}, {J{\'o}hannesson}, {Johnson}, {Johnson},
  {Johnson}, {Kamae}, {Katagiri}, {Kataoka}, {Kerr}, {Kn{\"o}dlseder}, {Kuss}, {Lande}, {Larsson}, {Latronico}, {Lavalley}, {Lemoine-Goumard}, {Longo}, {Loparco}, {Lott}, {Lovellette}, {Lubrano}, {Mazziotta}, {McConville}, {McEnery}, {Mehault}, {Michelson}, {Mitthumsiri}, {Mizuno}, {Moiseev}, {Monte}, {Monzani}, {Morselli}, {Moskalenko}, {Murgia}, {Naumann-Godo}, {Nemmen}, {Nishino}, {Norris}, {Nuss}, {Ohno}, {Ohsugi}, {Okumura}, {Omodei}, {Orienti}, {Orlando}, {Ormes}, {Paneque}, {Panetta}, {Perkins}, {Pesce-Rollins}, {Pierbattista}, {Piron}, {Pivato}, {Porter}, {Racusin}, {Rain{\`o}}, {Rando}, {Razzano}, {Razzaque}, {Reimer}, {Reimer}, {Reposeur}, {Reyes}, {Ritz}, {Rochester}, {Romoli}, {Roth}, {Sadrozinski}, {Sanchez}, {Saz Parkinson}, {Sbarra}, {Scargle}, {Sgr{\`o}}, {Siegal-Gaskins}, {Siskind}, {Spandre}, {Spinelli}, {Stephens}, {Suson}, {Tajima}, {Takahashi}, {Tanaka}, {Thayer}, {Thayer}, {Thompson}, {Tibaldo}, {Tinivella}, {Tosti}, {Troja}, {Usher}, {Vandenbroucke}, {Van Klaveren}, {Vasileiou},
  {Vianello}, {Vitale}, {Waite}, {Wallace}, {Winer}, {Wood}, {Wood}, {Wood}, {Yang}, \& {Zimmer}}]{Ackermann+12_Fermi}
{Ackermann}, M., {Ajello}, M., {Albert}, A., {et~al.} 2012{\natexlab{a}}, \apjs, 203, 4

\bibitem[{{Ackermann} {et~al.}(2012{\natexlab{b}}){Ackermann}, {Ajello}, {Atwood}, {Baldini}, {Ballet}, {Barbiellini}, {Bastieri}, {Bechtol}, {Bellazzini}, {Berenji}, {Blandford}, {Bloom}, {Bonamente}, {Borgland}, {Brandt}, {Bregeon}, {Brigida}, {Bruel}, {Buehler}, {Buson}, {Caliandro}, {Cameron}, {Caraveo}, {Cavazzuti}, {Cecchi}, {Charles}, {Chekhtman}, {Chiang}, {Ciprini}, {Claus}, {Cohen-Tanugi}, {Conrad}, {Cutini}, {de Angelis}, {de Palma}, {Dermer}, {Digel}, {Silva}, {Drell}, {Drlica-Wagner}, {Falletti}, {Favuzzi}, {Fegan}, {Ferrara}, {Focke}, {Fortin}, {Fukazawa}, {Funk}, {Fusco}, {Gaggero}, {Gargano}, {Germani}, {Giglietto}, {Giordano}, {Giroletti}, {Glanzman}, {Godfrey}, {Grove}, {Guiriec}, {Gustafsson}, {Hadasch}, {Hanabata}, {Harding}, {Hayashida}, {Hays}, {Horan}, {Hou}, {Hughes}, {J{\'o}hannesson}, {Johnson}, {Johnson}, {Kamae}, {Katagiri}, {Kataoka}, {Kn{\"o}dlseder}, {Kuss}, {Lande}, {Latronico}, {Lee}, {Lemoine-Goumard}, {Longo}, {Loparco}, {Lott}, {Lovellette}, {Lubrano}, {Mazziotta},
  {McEnery}, {Michelson}, {Mitthumsiri}, {Mizuno}, {Monte}, {Monzani}, {Morselli}, {Moskalenko}, {Murgia}, {Naumann-Godo}, {Norris}, {Nuss}, {Ohsugi}, {Okumura}, {Omodei}, {Orlando}, {Ormes}, {Paneque}, {Panetta}, {Parent}, {Pesce-Rollins}, {Pierbattista}, {Piron}, {Pivato}, {Porter}, {Rain{\`o}}, {Rando}, {Razzano}, {Razzaque}, {Reimer}, {Reimer}, {Sadrozinski}, {Sgr{\`o}}, {Siskind}, {Spandre}, {Spinelli}, {Strong}, {Suson}, {Takahashi}, {Tanaka}, {Thayer}, {Thayer}, {Thompson}, {Tibaldo}, {Tinivella}, {Torres}, {Tosti}, {Troja}, {Usher}, {Vandenbroucke}, {Vasileiou}, {Vianello}, {Vitale}, {Waite}, {Wang}, {Winer}, {Wood}, {Wood}, {Yang}, {Ziegler}, \& {Zimmer}}]{ackermann2012_DGE}
{Ackermann}, M., {Ajello}, M., {Atwood}, W.~B., {et~al.} 2012{\natexlab{b}}, \apj, 750, 3

\bibitem[{Ackermann {et~al.}(2011)Ackermann, Ajello, Baldini, Ballet, Barbiellini, Bastieri, Bechtol, Berenji, Bloom, Borgland, Bouvier, Buehler, Baldini, Bellazzini, Bregeon, Brez, Ballet, Barbiellini, Bastieri, Buson, Bonamente, Brigida, Bruel, Caliandro, Casandjian, Cecchi, Charles, Chekhtman, Chiang, Ciprini, Claus, Cohen-Tanugi, Conrad, Cutini, de~Angelis, de~Palma, Dermer, Digel, do~Couto~e Silva, Drell, Drlica-Wagner, Falletti, Favuzzi, Fegan, Ferrara, Fukazawa, Funk, Fusco, Gargano, Gasparrini, Gehrels, Germani, Giglietto, Giordano, Giroletti, Glanzman, Godfrey, Grenier, Guiriec, Gustafsson, Hadasch, Hayashida, Hays, Hughes, Jeltema, Johannesson, Johnson, Johnson, Kamae, Katagiri, Kataoka, Knödlseder, Kuss, Lande, Latronico, Lionetto, Garde, Longo, Loparco, Lott, Lovellette, Lubrano, Madejski, Mazziotta, McEnery, Mehault, Michelson, Mitthumsiri, Mizuno, Monte, Monzani, Morselli, Moskalenko, Murgia, Naumann-Godo, Norris, Nuss, Ohsugi, Okumura, Omodei, Orlando, Ormes, Ozaki, Paneque, Parent,
  Pesce-Rollins, Piron, Pivato, Porter, Profumo, Raino, Razzano, Reimer, Reimer, Ritz, Sadrozinski, Sbarra, Scargle, Schalk, Sgro, Siskind, Spandre, Spinelli, Strigari, Suson, Tajima, Takahashi, Tanaka, Thayer, Thayer, Thompson, Tibaldo, Tinivella, Torres, Troja, Uchiyama, Vandenbroucke, Vasileiou, Vianello, Vitale, Waite, Wang, Winer, Wood, Wood, Yang, \& Zimmer}]{ackermann2011fermiPWN}
Ackermann, M., Ajello, M., Baldini, L., {et~al.} 2011, \apj, 726, 35

\bibitem[{Ajello {et~al.}(2021)Ajello, Atwood, Axelsson, Bagagli, Bagni, Baldini, Bastieri, Bellardi, Bellazzini, Bissaldi, Bloom, Bonino, Bregeon, Brez, Bruel, Buehler, Buson, Cameron, Caraveo, Cavazzuti, Ceccanti, Chen, Cheung, Ciprini, Cognard, Cohen-Tanugi, Cutini, D’Ammando, de~la Torre~Luque, de~Palma, Digel, Dirirsa, Di~Lalla, Di~Venere, Domínguez, Fabiani, Ferrara, Fiori, Foglia, Fukazawa, Fusco, Gargano, Gasparrini, Giroletti, Glanzman, Green, Griffin, Grondin, Grove, Guillemot, Guiriec, Gustafsson, Hays, Horan, Jóhannesson, Johnson, Kamae, Kerr, Kuss, Larsson, Latronico, Lemoine-Goumard, Li, Liodakis, Longo, Loparco, Lovellette, Lubrano, Maldera, Manfreda, Martí-Devesa, Mazziotta, Menon, Mereu, Meyer, Michelson, Minuti, Mitthumsiri, Mizuno, Mongelli, Monzani, Moskalenko, Negro, Nuss, Ojha, Orienti, Orlando, Paccagnella, Paliya, Paneque, Pei, Perkins, Pesce-Rollins, Pinchera, Piron, Poon, Porter, Primavera, Principe, Racusin, Rainò, Rando, Rani, Rapposelli, Razzano, Razzaque, Reimer, Reimer,
  Russell, Saggini, Saz~Parkinson, Scolieri, Serini, Sgrò, Siskind, Smith, Spandre, Spinelli, Suson, Tajima, Thayer, Thompson, Tibaldo, Torres, Tosti, Valverde, Vigiani, \& Zaharijas}]{Ajello_2021}
Ajello, M., Atwood, W.~B., Axelsson, M., {et~al.} 2021, The Astrophysical Journal Supplement Series, 256, 12

\bibitem[{{An}(2019)}]{an2019}
{An}, H. 2019, \apj, 876, 150

\bibitem[{{Atwood} {et~al.}(2009){Atwood}, {Abdo}, {Ackermann}, {Althouse}, {Anderson}, {Axelsson}, {Baldini}, {Ballet}, {Band}, {Barbiellini}, {Bartelt}, {Bastieri}, {Baughman}, {Bechtol}, {B{\'e}d{\'e}r{\`e}de}, {Bellardi}, {Bellazzini}, {Berenji}, {Bignami}, {Bisello}, {Bissaldi}, {Blandford}, {Bloom}, {Bogart}, {Bonamente}, {Bonnell}, {Borgland}, {Bouvier}, {Bregeon}, {Brez}, {Brigida}, {Bruel}, {Burnett}, {Busetto}, {Caliandro}, {Cameron}, {Caraveo}, {Carius}, {Carlson}, {Casandjian}, {Cavazzuti}, {Ceccanti}, {Cecchi}, {Charles}, {Chekhtman}, {Cheung}, {Chiang}, {Chipaux}, {Cillis}, {Ciprini}, {Claus}, {Cohen-Tanugi}, {Condamoor}, {Conrad}, {Corbet}, {Corucci}, {Costamante}, {Cutini}, {Davis}, {Decotigny}, {DeKlotz}, {Dermer}, {de Angelis}, {Digel}, {do Couto e Silva}, {Drell}, {Dubois}, {Dumora}, {Edmonds}, {Fabiani}, {Farnier}, {Favuzzi}, {Flath}, {Fleury}, {Focke}, {Funk}, {Fusco}, {Gargano}, {Gasparrini}, {Gehrels}, {Gentit}, {Germani}, {Giebels}, {Giglietto}, {Giommi}, {Giordano}, {Glanzman},
  {Godfrey}, {Grenier}, {Grondin}, {Grove}, {Guillemot}, {Guiriec}, {Haller}, {Harding}, {Hart}, {Hays}, {Healey}, {Hirayama}, {Hjalmarsdotter}, {Horn}, {Hughes}, {J{\'o}hannesson}, {Johansson}, {Johnson}, {Johnson}, {Johnson}, {Johnson}, {Kamae}, {Katagiri}, {Kataoka}, {Kavelaars}, {Kawai}, {Kelly}, {Kerr}, {Klamra}, {Kn{\"o}dlseder}, {Kocian}, {Komin}, {Kuehn}, {Kuss}, {Landriu}, {Latronico}, {Lee}, {Lee}, {Lemoine-Goumard}, {Lionetto}, {Longo}, {Loparco}, {Lott}, {Lovellette}, {Lubrano}, {Madejski}, {Makeev}, {Marangelli}, {Massai}, {Mazziotta}, {McEnery}, {Menon}, {Meurer}, {Michelson}, {Minuti}, {Mirizzi}, {Mitthumsiri}, {Mizuno}, {Moiseev}, {Monte}, {Monzani}, {Moretti}, {Morselli}, {Moskalenko}, {Murgia}, {Nakamori}, {Nishino}, {Nolan}, {Norris}, {Nuss}, {Ohno}, {Ohsugi}, {Omodei}, {Orlando}, {Ormes}, {Paccagnella}, {Paneque}, {Panetta}, {Parent}, {Pearce}, {Pepe}, {Perazzo}, {Pesce-Rollins}, {Picozza}, {Pieri}, {Pinchera}, {Piron}, {Porter}, {Poupard}, {Rain{\`o}}, {Rando}, {Rapposelli}, {Razzano},
  {Reimer}, {Reimer}, {Reposeur}, {Reyes}, {Ritz}, {Rochester}, {Rodriguez}, {Romani}, {Roth}, {Russell}, {Ryde}, {Sabatini}, {Sadrozinski}, {Sanchez}, {Sander}, {Sapozhnikov}, {Parkinson}, {Scargle}, {Schalk}, {Scolieri}, {Sgr{\`o}}, {Share}, {Shaw}, {Shimokawabe}, {Shrader}, {Sierpowska-Bartosik}, {Siskind}, {Smith}, {Smith}, {Spandre}, {Spinelli}, {Starck}, {Stephens}, {Strickman}, {Strong}, {Suson}, {Tajima}, {Takahashi}, {Takahashi}, {Tanaka}, {Tenze}, {Tether}, {Thayer}, {Thayer}, {Thompson}, {Tibaldo}, {Tibolla}, {Torres}, {Tosti}, {Tramacere}, {Turri}, {Usher}, {Vilchez}, {Vitale}, {Wang}, {Watters}, {Winer}, {Wood}, {Ylinen}, \& {Ziegler}}]{atwood2009}
{Atwood}, W.~B., {Abdo}, A.~A., {Ackermann}, M., {et~al.} 2009, \apj, 697, 1071

\bibitem[{{Atwood} {et~al.}(2013){Atwood}, {Baldini}, {Bregeon}, {Bruel}, {Chekhtman}, {Cohen-Tanugi}, {Drlica-Wagner}, {Granot}, {Longo}, {Omodei}, {Pesce-Rollins}, {Razzaque}, {Rochester}, {Sgr{\`o}}, {Tinivella}, {Usher}, \& {Zimmer}}]{atwood2013}
{Atwood}, W.~B., {Baldini}, L., {Bregeon}, J., {et~al.} 2013, \apj, 774, 76

\bibitem[{{Ballet} {et~al.}(2023){Ballet}, {Bruel}, {Burnett}, {Lott}, \& {The Fermi-LAT collaboration}}]{Ballet+23}
{Ballet}, J., {Bruel}, P., {Burnett}, T.~H., {Lott}, B., \& {The Fermi-LAT collaboration}. 2023, arXiv e-prints, arXiv:2307.12546

\bibitem[{{Bhalerao} {et~al.}(2015){Bhalerao}, {Park}, {Dewey}, {Hughes}, {Mori}, \& {Lee}}]{bhalerao2015}
{Bhalerao}, J., {Park}, S., {Dewey}, D., {et~al.} 2015, \apj, 800, 65

\bibitem[{{Bhalerao} {et~al.}(2019){Bhalerao}, {Park}, {Schenck}, {Post}, \& {Hughes}}]{bhalerao2019}
{Bhalerao}, J., {Park}, S., {Schenck}, A., {Post}, S., \& {Hughes}, J.~P. 2019, \apj, 872, 31

\bibitem[{{Camilo} {et~al.}(2002{\natexlab{a}}){Camilo}, {Lorimer}, {Bhat}, {Gotthelf}, {Halpern}, {Wang}, {Lu}, \& {Mirabal}}]{Camilo2002G54}
{Camilo}, F., {Lorimer}, D.~R., {Bhat}, N.~D.~R., {et~al.} 2002{\natexlab{a}}, \apjl, 574, L71

\bibitem[{{Camilo} {et~al.}(2002{\natexlab{b}}){Camilo}, {Manchester}, {Gaensler}, {Lorimer}, \& {Sarkissian}}]{Camilo2002G292}
{Camilo}, F., {Manchester}, R.~N., {Gaensler}, B.~M., {Lorimer}, D.~R., \& {Sarkissian}, J. 2002{\natexlab{b}}, \apjl, 567, L71

\bibitem[{{Cotton} {et~al.}(2024){Cotton}, {Kothes}, {Camilo}, {Chandra}, {Buchner}, \& {Nyamai}}]{cotton2024}
{Cotton}, W.~D., {Kothes}, R., {Camilo}, F., {et~al.} 2024, \apjs, 270, 21

\bibitem[{de~Jager {et~al.}(2009)de~Jager, Ferreira, Djannati-Ataï, Dalton, Deil, Kosack, Renaud, Schwanke, \& Tibolla}]{dejager2009PWNe}
de~Jager, O.~C., Ferreira, S. E.~S., Djannati-Ataï, A., {et~al.} 2009, Unidentified Gamma-Ray Sources as Ancient Pulsar Wind Nebulae

\bibitem[{{Fesen} {et~al.}(2008){Fesen}, {Rudie}, {Hurford}, \& {Soto}}]{fesen2008}
{Fesen}, R., {Rudie}, G., {Hurford}, A., \& {Soto}, A. 2008, \apjs, 174, 379

\bibitem[{{Gaensler} \& {Slane}(2006)}]{gaenslerSlane2006}
{Gaensler}, B.~M. \& {Slane}, P.~O. 2006, \araa, 44, 17

\bibitem[{{Gaensler} \& {Wallace}(2003)}]{gaensler2003}
{Gaensler}, B.~M. \& {Wallace}, B.~J. 2003, \apj, 594, 326

\bibitem[{Ge {et~al.}(2020)Ge, Yuan, Lu, Tong, Zhou, Yan, Wang, Tuo, Li, \& Song}]{Ge_2020}
Ge, M.~Y., Yuan, J.~P., Lu, F.~J., {et~al.} 2020, The Astrophysical Journal Letters, 900, L7

\bibitem[{{Gelfand} {et~al.}(2015){Gelfand}, {Slane}, \& {Temim}}]{gelfand2015}
{Gelfand}, J.~D., {Slane}, P.~O., \& {Temim}, T. 2015, \apj, 807, 30

\bibitem[{{Grondin} {et~al.}(2013){Grondin}, {Romani}, {Lemoine-Goumard}, {Guillemot}, {Harding}, \& {Reposeur}}]{grondin2013}
{Grondin}, M.~H., {Romani}, R.~W., {Lemoine-Goumard}, M., {et~al.} 2013, \apj, 774, 110

\bibitem[{{H.~E.~S.~S. Collaboration} {et~al.}(2018){H.~E.~S.~S. Collaboration}, {Abdalla}, {Abramowski}, {Aharonian}, {Ait Benkhali}, {Akhperjanian}, {Andersson}, {Ang{\"u}ner}, {Arrieta}, {Aubert}, {Backes}, {Balzer}, {Barnard}, {Becherini}, {Becker Tjus}, {Berge}, {Bernhard}, {Bernl{\"o}hr}, {Blackwell}, {B{\"o}ttcher}, {Boisson}, {Bolmont}, {Bordas}, {Bregeon}, {Brun}, {Brun}, {Bryan}, {Bulik}, {Capasso}, {Carr}, {Carrigan}, {Casanova}, {Cerruti}, {Chakraborty}, {Chalme-Calvet}, {Chaves}, {Chen}, {Chevalier}, {Chr{\'e}tien}, {Colafrancesco}, {Cologna}, {Condon}, {Conrad}, {Couturier}, {Cui}, {Davids}, {Degrange}, {Deil}, {Devin}, {deWilt}, {Dirson}, {Djannati-Ata{\"\i}}, {Domainko}, {Donath}, {Drury}, {Dubus}, {Dutson}, {Dyks}, {Edwards}, {Egberts}, {Eger}, {Ernenwein}, {Eschbach}, {Farnier}, {Fegan}, {Fernandes}, {Fiasson}, {Fontaine}, {F{\"o}rster}, {Funk}, {F{\"u}{\ss}ling}, {Gabici}, {Gajdus}, {Gallant}, {Garrigoux}, {Giavitto}, {Giebels}, {Glicenstein}, {Gottschall}, {Goyal}, {Grondin}, {Hadasch},
  {Hahn}, {Haupt}, {Hawkes}, {Heinzelmann}, {Henri}, {Hermann}, {Hervet}, {Hillert}, {Hinton}, {Hofmann}, {Hoischen}, {Holler}, {Horns}, {Ivascenko}, {Jacholkowska}, {Jamrozy}, {Janiak}, {Jankowsky}, {Jankowsky}, {Jingo}, {Jogler}, {Jouvin}, {Jung-Richardt}, {Kastendieck}, {Katarzy{\'n}ski}, {Katz}, {Kerszberg}, {Kh{\'e}lifi}, {Kieffer}, {King}, {Klepser}, {Klochkov}, {Klu{\'z}niak}, {Kolitzus}, {Komin}, {Kosack}, {Krakau}, {Kraus}, {Krayzel}, {Kr{\"u}ger}, {Laffon}, {Lamanna}, {Lau}, {Lees}, {Lefaucheur}, {Lefranc}, {Lemi{\`e}re}, {Lemoine-Goumard}, {Lenain}, {Leser}, {Lohse}, {Lorentz}, {Liu}, {L{\'o}pez-Coto}, {Lypova}, {Marandon}, {Marcowith}, {Mariaud}, {Marx}, {Maurin}, {Maxted}, {Mayer}, {Meintjes}, {Meyer}, {Mitchell}, {Moderski}, {Mohamed}, {Mohrmann}, {Mor{\r{a}}}, {Moulin}, {Murach}, {de Naurois}, {Niederwanger}, {Niemiec}, {Oakes}, {O'Brien}, {Odaka}, {{\"O}ttl}, {Ohm}, {de O{\~n}a Wilhelmi}, {Ostrowski}, {Oya}, {Padovani}, {Panter}, {Parsons}, {Paz Arribas}, {Pekeur}, {Pelletier}, {Perennes},
  {Petrucci}, {Peyaud}, {Pita}, {Poon}, {Prokhorov}, {Prokoph}, {P{\"u}hlhofer}, {Punch}, {Quirrenbach}, {Raab}, {Reimer}, {Reimer}, {Renaud}, {de los Reyes}, {Rieger}, {Romoli}, {Rosier-Lees}, {Rowell}, {Rudak}, {Rulten}, {Sahakian}, {Salek}, {Sanchez}, {Santangelo}, {Sasaki}, {Schlickeiser}, {Sch{\"u}ssler}, {Schulz}, \& {Schwanke}}]{2018_HESS_PWN_Pop}
{H.~E.~S.~S. Collaboration}, {Abdalla}, H., {Abramowski}, A., {et~al.} 2018, \aap, 612, A2

\bibitem[{{Hughes} {et~al.}(2001){Hughes}, {Slane}, {Burrows}, {Garmire}, {Nousek}, {Olbert}, \& {Keohane}}]{hughes2001}
{Hughes}, J.~P., {Slane}, P.~O., {Burrows}, D.~N., {et~al.} 2001, \apjl, 559, L153

\bibitem[{{Hughes} {et~al.}(2003){Hughes}, {Slane}, {Park}, {Roming}, \& {Burrows}}]{hughes2003}
{Hughes}, J.~P., {Slane}, P.~O., {Park}, S., {Roming}, P. W.~A., \& {Burrows}, D.~N. 2003, \apjl, 591, L139

\bibitem[{{Hurley-Walker} {et~al.}(2017){Hurley-Walker}, {Callingham}, {Hancock}, {Franzen}, {Hindson}, {Kapi{\'n}ska}, {Morgan}, {Offringa}, {Wayth}, {Wu}, {Zheng}, {Murphy}, {Bell}, {Dwarakanath}, {For}, {Gaensler}, {Johnston-Hollitt}, {Lenc}, {Procopio}, {Staveley-Smith}, {Ekers}, {Bowman}, {Briggs}, {Cappallo}, {Deshpande}, {Greenhill}, {Hazelton}, {Kaplan}, {Lonsdale}, {McWhirter}, {Mitchell}, {Morales}, {Morgan}, {Oberoi}, {Ord}, {Prabu}, {Shankar}, {Srivani}, {Subrahmanyan}, {Tingay}, {Webster}, {Williams}, \& {Williams}}]{hurley-walker2017}
{Hurley-Walker}, N., {Callingham}, J.~R., {Hancock}, P.~J., {et~al.} 2017, \mnras, 464, 1146

\bibitem[{Kothes(2013)}]{kothes2013}
Kothes, R. 2013, \apj, 765, 149

\bibitem[{Lange {et~al.}(2025)Lange, Eagle, Kargaltsev, Kuiper, \& Hare}]{lange2025}
Lange, A., Eagle, J., Kargaltsev, O., Kuiper, L., \& Hare, J. 2025, \apj [\eprint[arXiv]{2506.16687}], accepted for publication

\bibitem[{{Leahy} {et~al.}(2008){Leahy}, {Tian}, \& {Wang}}]{Leahy2008G54}
{Leahy}, D.~A., {Tian}, W., \& {Wang}, Q.~D. 2008, \aj, 136, 1477

\bibitem[{{Lee} {et~al.}(2009){Lee}, {Koo}, {Moon}, {Sakon}, {Onaka}, {Jeong}, {Kaneda}, {Nozawa}, \& {Kozasa}}]{Lee2009}
{Lee}, H.-G., {Koo}, B.-C., {Moon}, D.-S., {et~al.} 2009, \apj, 706, 441

\bibitem[{{Lee} {et~al.}(2010){Lee}, {Park}, {Hughes}, {Slane}, {Gaensler}, {Ghavamian}, \& {Burrows}}]{lee2010}
{Lee}, J.-J., {Park}, S., {Hughes}, J.~P., {et~al.} 2010, \apj, 711, 861

\bibitem[{{Liu} {et~al.}(2024){Liu}, {Zeng}, {Xin}, {Liu}, \& {Zhang}}]{liu2024}
{Liu}, Y.-M., {Zeng}, H.-D., {Xin}, Y.-L., {Liu}, S.-M., \& {Zhang}, Y. 2024, Research in Astronomy and Astrophysics, 24, 075016

\bibitem[{Long {et~al.}(2022)Long, Patnaude, Plucinsky, \& Gaetz}]{Long2022G292}
Long, X., Patnaude, D.~J., Plucinsky, P.~P., \& Gaetz, T.~J. 2022, \apj, 936, 24

\bibitem[{{Martin} {et~al.}(2014){Martin}, {Torres}, {Cillis}, \& {de O{\~n}a Wilhelmi}}]{martin2014}
{Martin}, J., {Torres}, D.~F., {Cillis}, A., \& {de O{\~n}a Wilhelmi}, E. 2014, \mnras, 443, 138

\bibitem[{{Mattana} {et~al.}(2009){Mattana}, {Falanga}, {G{\"o}tz}, {Terrier}, {Esposito}, {Pellizzoni}, {De Luca}, {Marandon}, {Goldwurm}, \& {Caraveo}}]{mattana2009}
{Mattana}, F., {Falanga}, M., {G{\"o}tz}, D., {et~al.} 2009, \apj, 694, 12

\bibitem[{{McConnell} {et~al.}(2020){McConnell}, {Hale}, {Lenc}, {Banfield}, {Heald}, {Hotan}, {Leung}, {Moss}, {Murphy}, {O'Brien}, {Pritchard}, {Raja}, {Sadler}, {Stewart}, {Thomson}, {Whiting}, {Allison}, {Amy}, {Anderson}, {Ball}, {Bannister}, {Bell}, {Bock}, {Bolton}, {Bunton}, {Chippendale}, {Collier}, {Cooray}, {Cornwell}, {Diamond}, {Edwards}, {Gupta}, {Hayman}, {Heywood}, {Jackson}, {Koribalski}, {Lee-Waddell}, {McClure-Griffiths}, {Ng}, {Norris}, {Phillips}, {Reynolds}, {Roxby}, {Schinckel}, {Shields}, {Tremblay}, {Tzioumis}, {Voronkov}, \& {Westmeier}}]{mcconnell2020}
{McConnell}, D., {Hale}, C.~L., {Lenc}, E., {et~al.} 2020, \pasa, 37, e048

\bibitem[{{Park} {et~al.}(2007){Park}, {Hughes}, {Slane}, {Burrows}, {Gaensler}, \& {Ghavamian}}]{park2007}
{Park}, S., {Hughes}, J.~P., {Slane}, P.~O., {et~al.} 2007, \apjl, 670, L121

\bibitem[{{Perley} \& {Butler}(2017)}]{perley17}
{Perley}, R.~A. \& {Butler}, B.~J. 2017, \apjs, 230, 7

\bibitem[{{Porter} {et~al.}(2006){Porter}, {Moskalenko}, \& {Strong}}]{porter2006}
{Porter}, T.~A., {Moskalenko}, I.~V., \& {Strong}, A.~W. 2006, \apjl, 648, L29

\bibitem[{{Rousseau} {et~al.}(2012){Rousseau}, {Grondin}, {Van Etten}, {Lemoine-Goumard}, {Bogdanov}, {Hessels}, {Kaspi}, {Arzoumanian}, {Camilo}, {Casandjian}, {Espinoza}, {Johnston}, {Lyne}, {Smith}, {Stappers}, \& {Caliandro}}]{2012PWNJ1857}
{Rousseau}, R., {Grondin}, M.~H., {Van Etten}, A., {et~al.} 2012, \aap, 544, A3

\bibitem[{Rudie {et~al.}(2007)Rudie, Hester, \& Allen}]{Rudie2007CrabNebulaAge}
Rudie, G.~C., Hester, J.~J., \& Allen, G.~E. 2007, \apj, 661, 417

\bibitem[{{Safi-Harb} \& {Gonzalez}(2002)}]{safiharb2002}
{Safi-Harb}, S. \& {Gonzalez}, M.~E. 2002, in Astronomical Society of the Pacific Conference Series, Vol. 262, The High Energy Universe at Sharp Focus: Chandra Science, ed. E.~M. {Schlegel} \& S.~D. {Vrtilek}, 315

\bibitem[{{Shibata} {et~al.}(2011){Shibata}, {Ishikawa}, \& {Sekiguchi}}]{Shibata+11}
{Shibata}, T., {Ishikawa}, T., \& {Sekiguchi}, S. 2011, \apj, 727, 38

\bibitem[{{Slane} {et~al.}(2008){Slane}, {Helfand}, {Reynolds}, {Gaensler}, {Lemiere}, \& {Wang}}]{slane2008}
{Slane}, P., {Helfand}, D.~J., {Reynolds}, S.~P., {et~al.} 2008, \apjl, 676, L33

\bibitem[{{Smith} {et~al.}(2023){Smith}, {Abdollahi}, {Ajello}, {Bailes}, {Baldini}, {Ballet}, {Baring}, {Bassa}, {Gonzalez}, {Bellazzini}, {Berretta}, {Bhattacharyya}, {Bissaldi}, {Bonino}, {Bottacini}, {Bregeon}, {Bruel}, {Burgay}, {Burnett}, {Cameron}, {Camilo}, {Caputo}, {Caraveo}, {Cavazzuti}, {Chiaro}, {Ciprini}, {Clark}, {Cognard}, {Corongiu}, {Orestano}, {Crnogorcevic}, {Cuoco}, {Cutini}, {D'Ammando}, {de Angelis}, {DeCesar}, {De Gaetano}, {de Menezes}, {Deneva}, {de Palma}, {Di Lalla}, {Dirirsa}, {Di Venere}, {Dom{\'\i}nguez}, {Dumora}, {Fegan}, {Ferrara}, {Fiori}, {Fleischhack}, {Flynn}, {Franckowiak}, {Freire}, {Fukazawa}, {Fusco}, {Galanti}, {Gammaldi}, {Gargano}, {Gasparrini}, {Giacchino}, {Giglietto}, {Giordano}, {Giroletti}, {Green}, {Grenier}, {Guillemot}, {Guiriec}, {Gustafsson}, {Harding}, {Hays}, {Hewitt}, {Horan}, {Hou}, {Jankowski}, {Johnson}, {Johnson}, {Johnston}, {Kataoka}, {Keith}, {Kerr}, {Kramer}, {Kuss}, {Latronico}, {Lee}, {Li}, {Li}, {Limyansky}, {Longo}, {Loparco}, {Lorusso},
  {Lovellette}, {Lower}, {Lubrano}, {Lyne}, {Maan}, {Maldera}, {Manchester}, {Manfreda}, {Marelli}, {Mart{\'\i}-Devesa}, {Mazziotta}, {McEnery}, {Mereu}, {Michelson}, {Mickaliger}, {Mitthumsiri}, {Mizuno}, {Moiseev}, {Monzani}, {Morselli}, {Negro}, {Nemmen}, {Nieder}, {Nuss}, {Omodei}, {Orienti}, {Orlando}, {Ormes}, {Palatiello}, {Paneque}, {Panzarini}, {Parthasarathy}, {Persic}, {Pesce-Rollins}, {Pillera}, {Poon}, {Porter}, {Possenti}, {Principe}, {Rain{\`o}}, {Rando}, {Ransom}, {Ray}, {Razzano}, {Razzaque}, {Reimer}, {Reimer}, {Renault-Tinacci}, {Romani}, {S{\'a}nchez-Conde}, {Parkinson}, {Scotton}, {Serini}, {Sgr{\`o}}, {Shannon}, {Sharma}, {Shen}, {Siskind}, {Spandre}, {Spinelli}, {Stappers}, {Stephens}, {Suson}, {Tabassum}, {Tajima}, {Tak}, {Theureau}, {Thompson}, {Tibolla}, {Torres}, {Valverde}, {Venter}, {Wadiasingh}, {Wang}, {Wang}, {Wang}, {Weltevrede}, {Wood}, {Yan}, {Zaharijas}, {Zhang}, \& {Zhu}}]{smith2023}
{Smith}, D.~A., {Abdollahi}, S., {Ajello}, M., {et~al.} 2023, \apj, 958, 191

\bibitem[{{Strong} {et~al.}(2000){Strong}, {Moskalenko}, \& {Reimer}}]{strong2000}
{Strong}, A.~W., {Moskalenko}, I.~V., \& {Reimer}, O. 2000, \apj, 537, 763

\bibitem[{{Swartz} {et~al.}(2015){Swartz}, {Pavlov}, {Clarke}, {Castelletti}, {Zavlin}, {Bucciantini}, {Karovska}, {van der Horst}, {Yukita}, \& {Weisskopf}}]{swartz2015}
{Swartz}, D.~A., {Pavlov}, G.~G., {Clarke}, T., {et~al.} 2015, \apj, 808, 84

\bibitem[{{Tanaka} \& {Takahara}(2013)}]{tanaka2013}
{Tanaka}, S.~J. \& {Takahara}, F. 2013, \mnras, 429, 2945

\bibitem[{{Temim} {et~al.}(2022){Temim}, {Slane}, {Raymond}, {Patnaude}, {Murray}, {Ghavamian}, {Renzo}, \& {Jacovich}}]{temim2022}
{Temim}, T., {Slane}, P., {Raymond}, J.~C., {et~al.} 2022, \apj, 932, 26

\bibitem[{{Winkler} {et~al.}(2009){Winkler}, {Twelker}, {Reith}, \& {Long}}]{winkler2009}
{Winkler}, P.~F., {Twelker}, K., {Reith}, C.~N., \& {Long}, K.~S. 2009, \apj, 692, 1489

\bibitem[{{Wood} {et~al.}(2021){Wood}, {Caputo}, {Charles}, {Di Mauro}, {Magill}, {Perkins}, \& {Fermi-LAT Collaboration}}]{Wood+17_fermipy}
{Wood}, M., {Caputo}, R., {Charles}, E., {et~al.} 2021, in International Cosmic Ray Conference, Vol. 301, ICRC2017, 824

\bibitem[{{Zabalza}(2015)}]{Zabalza+15}
{Zabalza}, V. 2015, in International Cosmic Ray Conference, Vol.~34, ICRC2015, 922

\bibitem[{{Zharikov} {et~al.}(2008){Zharikov}, {Shibanov}, {Zyuzin}, {Mennickent}, \& {Komarova}}]{zharikov2008}
{Zharikov}, S.~V., {Shibanov}, Y.~A., {Zyuzin}, D.~A., {Mennickent}, R.~E., \& {Komarova}, V.~N. 2008, \aap, 492, 805

\bibitem[{{Zharikov} {et~al.}(2013){Zharikov}, {Zyuzin}, {Shibanov}, \& {Mennickent}}]{zharikov2013}
{Zharikov}, S.~V., {Zyuzin}, D.~A., {Shibanov}, Y.~A., \& {Mennickent}, R.~E. 2013, \aap, 554, A120

\bibitem[{{Zhu} {et~al.}(2018){Zhu}, {Zhang}, \& {Fang}}]{zhu2018}
{Zhu}, B.-T., {Zhang}, L., \& {Fang}, J. 2018, \aap, 609, A110

\bibitem[{{Zyuzin} {et~al.}(2009){Zyuzin}, {Danilenko}, {Zharikov}, \& {Shibanov}}]{zyuzin2009}
{Zyuzin}, D.~A., {Danilenko}, A.~A., {Zharikov}, S.~V., \& {Shibanov}, Y.~A. 2009, \aap, 508, 855

\end{thebibliography}

\end{document}